\documentclass[prd,aps,a4paper,floatfix,amsmath,amssymb,twocolumn,nofootinbib]{revtex4-1}
\usepackage{amssymb,amsmath,amsthm,xcolor,graphicx}
\usepackage[margin=2cm]{geometry}
\usepackage[pdftex,breaklinks,colorlinks,
linkcolor=blue,
citecolor=teal,
anchorcolor=red,
urlcolor=cyan]{hyperref}
\newcommand{\new}{}

\begin{document}

%%%%%%%%%%%%%%%%%%%%%%%%%%%%%%%%%%%%%%%%%%%%%%%%%%%%%%%%%%%%%%%%%%%%%%%%%%%%%

\title{Simulations of gravitational collapse in null coordinates IV:
  evolving through the event horizon,
  with an application to the spherical charged scalar field}
\author{Carsten Gundlach and Laetitia Martel}
\affiliation{Mathematical Sciences, University of Southampton,
  Southampton SO17 1BJ, United Kingdom} 
\date{18 November 2025, revised 9 January 2026}

%%%%%%%%%%%%%%%%%%%%%%%%%%%%%%%%%%%%%%%%%%%%%%%%%%%%%%%%%%%%%%%%%%%%%%%%%%%%%

\begin{abstract}
We consider line elements of the form $-2G\,du\,(dx+B\,du) +
R^2(...)$, where $(...)$ does not contain $dx$. Surfaces of constant
$u$ are then null surfaces, and their affinely parameterised
generators have tangent vector $G^{-1}\partial_x$. Considering $u$ as
the time coordinate, we can evolve either $R$ or $G$, with the other
one found by solving the Raychaudhuri equation along the null
generators, or we can evolve both. This choice of {\em formulation} is
independent from the remaining {\em gauge} choice $x\to x'(u,x,...)$
in the line element above, which is fixed incrementally by the choice
of $B$. For example, we can evolve $G$, in order to be able to evolve
through an event horizon, and use $B$ to adapt the coordinates to
type-II critical collapse. As a demonstration of these ideas, we
consider a charged scalar field in spherical symmetry. We consider two
settings: a domain where the outgoing null cones emanate from a
regular centre $R=0$, and a domain where they emanate from an
ingoing-null boundary. In both settings, we demonstrate convergence
with resolution, within each formulation and between the three
formulations. As testbeds, we compute the critical exponents and
periodic fine-structures of the black hole charge and mass scaling
laws in a one-parameter family of charged regular initial data, and
examples of perturbed extremal Reissner-Nordstr\"om solutions.

\end{abstract}

%%%%%%%%%%%%%%%%%%%%%%%%%%%r%%%%%%%%%%%%%%%%%%%%%%%%%%%%%%%%%%%%%%%%%%%%%%%%%%

\maketitle

\tableofcontents

%%%%%%%%%%%%%%%%%%%%%%%%%%%%%%%%%%%%%%%%%%%%%%%%%%%%%%%%%%%%%%%%%%%%%%%%%%%%%%%

\section{Introduction}

%%%%%%%%%%%%%%%%%%%%%%%%%%%%%%%%%%%%%%%%%%%%%%%%%%%%%%%%%%%%%%%%%%%%%%%%%%%%%%%

\subsection{Overview}

%%%%%%%%%%%%%%%%%%%%%%%%%%%%%%%%%%%%%%%%%%%%%%%%%%%%%%%%%%%%%%%%%%%%%%%%%%%%%%%

In this series of papers, starting with \cite{axinull_formulation}, we
have been developing numerical simulations of gravitational collapse
on null cones emanating from a regular centre. {\new Null coordinates are an
  established tool in spherical symmetry, but our main motivation is
  simulating critical collapse on null cones beyond spherical
  symmetry. As long as no caustics form, null coordinates are ideal
  for type-II critical collapse because they allow us to adapt
  the coordinate system to the approximate self-similarity and large
  range of scales seen there.

In the present paper, we introduce new formulations of the Einstein
equations in null coordinates that allow us to evolve from regular
data through the black hole event horizon, while keeping the radial
coordinate arbitrary. The motivation for this will be given in more
detail in the next Subsection, but, in brief, we want to evolve through
horizons to accurately characterise black holes, while we also want to keep
the radial coordinate arbitrary in order to adapt the coordinates to
type-II critical collapse.

To make the presentation more concrete, after this Introduction we
apply the new formulations to the spherical Einstein-Maxwell-charged
scalar field system. After demonstrating convergence, we turn to type-II
critical collapse in this system, and obtain some new physics
results. In particular, we verify the critical exponents for
the black hole mass and charge, calculate the mass and charge fine
structures, and present preliminary evidence that the charge fine
structure is not universal. We plan to return to non-spherical type-II
critical collapse elsewhere.

The spherical Einstein-Maxwell-charged scalar systems is of current
interest also because of the stable extremal critical collapse conjecture of
Kehle and Unger \cite{KehleUnger22,KehleUnger24}. This says that
there is an open region of solution space in which the threshold of
black hole formation consists of extremal black holes.

In this context, we will need to evolve initial data posed on an
outgoing and an ingoing null cone that meet at 2-sphere. The domain of
dependence of these data is then a ``null rectangle'' in the
symmetry-reduced spacetime. We present an algorithm for this setup,
and demonstrate convergence in selected example solutions. We plan to
investigate extremal critical collapse in this system elsewhere.

}

%%%%%%%%%%%%%%%%%%%%%%%%%%%%%%%%%%%%%%%%%%%%%%%%%%%%%%%%%%%%%%%%%%%%%%%%%%%%%%%

\subsection{The problem posed by evolving through an event horizon in
  null coordinates}

%%%%%%%%%%%%%%%%%%%%%%%%%%%%%%%%%%%%%%%%%%%%%%%%%%%%%%%%%%%%%%%%%%%%%%%%%%%%%%%

In Paper~II of the series \cite{axinull_critscalar}, we investigated
the threshold of black hole formation for a massless scalar field
coupled to general relativity in axisymmetry, adapting the coordinates
to the expected self-similar critical solution. While, at moderate
non-sphericity, this allowed us to fine-tune to the threshold of black
hole formation down to machine precision, without any need for adaptive
mesh-refinement, we could not reach non-sphericities as high as others
had reached before us. In hindsight, this was due both to our choice
of radial gauge (how the radial coordinate $x$ labels points on each
null cone generator) and our choice of formulation (which of the
Einstein equations are used in the code and which are not). For our
notation for the metric, see already Eq.~(\ref{genmetric}) below.

Our {\em gauge} choice in Paper~II was to make the area radius $R$ a
function $R(u,x)$ of retarded time $u$ and radial coordinate $x$ only,
but not the two angular coordinates: we called this local shifted
Bondi (lsB) gauge, as it is a generalisation of Bondi gauge $R=x$. We noted
that near the event horizon, in highly non-spherical spacetimes, this
was not a good gauge choice, as it caused the radial shift $B$ required to keep
$R$ independent of the angles to diverge.

Separately, we chose the {\em formulation} where the metric
coefficient $G$ is obtained by solving the Raychaudhuri equation along
each outgoing null coordinate line. This formulation requires the area
radius $R$ to be strictly increasing along each null cone generator,
that is $R_{,x}>0$. In spherical symmetry, this still allows our time
slices (outgoing null cones) to approach the event horizon
asymptotically. Beyond spherical symmetry, there will be retarded
times $u_0$ and $u_1$ such that the full event horizon is seen only
after $u=u_1$, but in a formulation that requires $R_{,x}>0$ the code
must stop already at $u=u_0$.

The combination of lsB gauge and the formulation where $G$ is
constrained meant in practice that the simulations of highly
non-spherical spacetimes near the black-hole treshold had to stop
before they could be classified as either collapsing or dispersing.

This paper presents a second formulation, where the Raychaudhuri
equation is solved for $R$, rather than for $G$. In the literature,
this approach has been used only together with the affine radial gauge
$G=1$. Here we evolve $G$, which allows us to use an arbitrary radial
shift $B$, and in particular one adapted to critical collapse. 

We also present a third formulation, where both $R$ and $G$ are
evolved and the Raychaudhuri equation is used only to construct
initial data. In the literature, this formulation has been used only
with double null gauge $B=0$, but $B$ can in fact be chosen freely.

{\new One of our main points is the observation} that all three
formulations can be combined with any radial gauge choice.

The gauge problem in critical collapse mentioned above does not arise
in spherical symmetry, and will be addressed when we return to
axisymmetry elsewhere.

%%%%%%%%%%%%%%%%%%%%%%%%%%%%%%%%%%%%%%%%%%%%%%%%%%%%%%%%%%%%%%%%%%%%%%%%%%%%%%%

\subsection{Formulations and radial gauges in null coordinates}
\label{section:formulations}

%%%%%%%%%%%%%%%%%%%%%%%%%%%%%%%%%%%%%%%%%%%%%%%%%%%%%%%%%%%%%%%%%%%%%%%%%%%%%%%

We now give a more precise statement of what we just said about
formulations. For clarity, we do not fix the spacetime
dimension or symmetry. (In the remainder of the paper, we will
restrict to spherical symmetry in 3+1 dimensions.) We consider line
elements of the form
\begin{eqnarray}
\label{genmetric}
ds^2&=&-2G\,du\,(dx+B\,du) \nonumber \\
&&+ R^2 \gamma_{ij}(d\theta^i+\beta^i\,du)(d\theta^j+\beta^j\,du).
\end{eqnarray}
(The quantity $B$ replaces the metric coefficient $H:=-g_{uu}$ that we
used in Papers I--III, with $B:=H/(2G)$. This is just for
convenience of notation.) The spacetime dimension is $n+2$, and the
$\theta^i$ are coordinates on an $n$-dimensional manifold, typically
$S^n$. From this form of $g_{\mu\nu}$ it follows that
$g^{uu}=g^{ui}=0$, and $g^{ux}=-G^{-1}$. Hence the surfaces of
constant $u$ are null surfaces, and their affinely parameterised
generators have tangent vector
\begin{equation}
U^a:=-\nabla^a u=G^{-1}(\partial_x)^a. 
\end{equation}
We must have $G>0$ for the metric to be regular. $R$ is defined by
restricting $\gamma_{ij}$ to have a fixed determinant (such as its
value on a unit sphere $S^n$ in standard coordinates $\theta^i$). The
remaining gauge freedom in this ansatz for the line element is to
re-label $u$ as $u\to u'(u)$, and to relabel the radial coordinate $x$
as $x\to x'(u,x,\theta^i)$. We call a choice of coordinate $x$ a
radial gauge.

The expansion of the congruence of null
generators can be expressed in terms of $R$ as
\begin{equation}
\label{defrhoplus}
\rho_+={d\ln R\over d\lambda},
\end{equation}
where we have defined
\begin{equation}
{d\over d\lambda}:=G^{-1}\partial_x.
\end{equation}
(See also Paper~I \cite{axinull_formulation} for the
definition of $\rho_+$ from first principles.) The quantity
\begin{equation}
\lambda(u,x,\theta^i):=\int_0^x G(u,x',\theta^i)\,dx'
\end{equation}
is a preferred affine parameter along the null generators. (We shall
refer to it as ``the'' affine parameter).

The $xx$-component of the trace-reversed Einstein
equations can be written as
\begin{equation}
\label{genericRxxeqn}
R_{,xx}-{G_{,x}\over G}R_{,x}+KR=0,
\end{equation}
or as
\begin{equation}
\label{genericGxeqn}
\left(\ln{G\over R_{,x}}\right)_{,x}={KR\over R_{,x}},
\end{equation}
or as
\begin{equation}
\label{genericd2Rdlambda}
{d^2R\over d\lambda^2}+{K\over G^2}R=0,
\end{equation}
or as the Raychaudhuri equation
\begin{equation}
\label{genericRaychaudhuri}
{d\rho_+\over d\lambda}=-\rho_+^2-{K\over G^2},
\end{equation}
where the term $K$ contains shear and matter terms, and is
non-negative in vacuum or with matter obeying the dominant energy
condition. 

From (\ref{genericd2Rdlambda}), with $K\ge 0$, we see that once
$dR/d\lambda$ has become negative, $R$ reaches zero in finite affine
parameter. Equivalently, from (\ref{genericRaychaudhuri}) we see that
once $\rho_+$ has become negative, it blows up in finite affine
parameter. We can now make the above statement precise: In a
non-spherical collapse spacetime, generically there will be $u_0<u_1$
such that for any coordinate null cone with $u_0<u<u_1$, the expansion
$\rho_+$ remains positive on some of its generators, while on others
it becomes negative and then blows up. This could be because the
generator has reached a curvature singularity, or because the null
cone has developed a caustic.

We now come to the role of the Raychaudhuri equation in numerical
relativity on null cones.  In its form (\ref{genericRxxeqn}) and with
$G$ given, it can be solved as a second-order ordinary differential
equation (from now on, ODE) in $x$ for $R$. In its form
(\ref{genericGxeqn}) and with $R$ given, it can be solved as a
first-order ODE for $G$. We also have wave equation-like evolution
equations $G_{,ux}=...$ and $R_{,ux}=...$ for both $G$ and $R$. This
gives rise to three formulations, each of which admits an arbitrary
radial gauge:
\begin{itemize}
\item evolve $R$ and constrain $G$ ({\bf
  evolve-$R$ formulation}, from now on {\bf eR});
\item evolve $G$ and constrain $R$ ({\bf
  evolve-$G$ formulation}, from now on {\bf eG});
\item evolve $R$ and $G$ freely and impose the constraint between them
  only at $u=0$ ({\bf free-evolution formulation}, from now on {\bf fe}).
\end{itemize}
The eR formulation requires $R_{,x}>0$, because we divide by
$R_{,x}$ in (\ref{genericGxeqn}), and so it breaks down
once any part of a coordinate null cone intersects the black hole
region. The other two formulations allow us to evolve further. It is
this observation that motivates us to consider eG and fe in the present paper. 

After the radial gauge has been fixed in the initial data at $u=0$,
typically by specifying either $R$ or $G$ as a function of
$(x,\theta^i)$, it is controlled {\em incrementally} by the choice of
the quantity $B$. The prototype incremental gauge choice is
double-null gauge $B=0$, which makes $x$ an ingoing null coordinate.

In addition to the three formulations above, there are two particular
{\em combinations} of formulation and gauge that remove the need for one
evolution equation, and which we feel should be characterised as formulations:
\begin{itemize}
\item set $R=x$ and constrain $G$ ({\bf Bondi formulation}); 
\item set $G=1$ and constrain $R$ ({\bf affine formulation}).
\end{itemize}
In both these formulations, all time evolution equations (equations
containing $u$-derivatives) correspond to physical (matter and
gravitational wave) degrees of freedom, so they are ``maximally
constrained'' in the language of numerical relativity. By contrast, eR
and eG have one gauge evolution equation, and fe has
two. Table~\ref{table:formulations} summarises the five formulations
we have just discussed. The observation that the choices of
formulation and gauge are independent is one of the main points of the
present paper.

One can retain some of the stability advantages of a maximally
constrained formulation in the eR formulation by restricting 
to the family of ``local shifted Bondi'' (lsB) radial gauges, where
$R=R(u,x)$. We did this in Paper~II. Similarly,
in the eG formulation, we can restrict to ``local shifted affine''
(lsa) gauges, where $G=G(u,x)$, to make it almost maximally constrained.

%%%%%%%%%%%%%%%%%%%%%%%%%%%%%%%%%%%%%%%%%%%%%%%%%%%%%%%%%%%%%%%%%%%%%%%%%%%%%%%
\begin{table*}
\setlength{\tabcolsep}{12pt} % Default value: 6pt
\renewcommand{\arraystretch}{1.3} % Default value: 1
\begin{tabular}{l||l|l|l|l}
Formulation & Initial gauge & Evolved gauge & Restricted to & Gauge evolution equations \\
\hline
Bondi & $R=x$ & $R=x$ &$R_{,x}>0$ & 0 \\
affine & $G=1$ & $G=1$ &--- & 0 \\
eR & any $R$ or $G$ & any $B$ & $R_{,x}>0$ & 1 \\
eG & any $R$ or $G$ & any $B$ & --- & 1 \\
free evolution & any $R$ or $G$ & any $B$ & --- & 2 \\
\end{tabular}
\caption{Summary of the five formulations discussed in
  Sec.~\ref{section:formulations}. ``Initial gauge'' denotes the
  freedom to label points on $u=0$ by the coordinate $x$, so ``any
  $R$'' is short for ``any choice of $R(0,x,\theta^i)$'' and similarly
  for ``any $G$''. ``Evolved gauge'' denotes how the coordinate $x$ is
  propagated to $u>0$, and ``any $B$'' is short for ``any choice of
  $B(u,x,\theta^i)$''. The restriction to $R_{,x}>0$ arises when the
  Raychaudhuri equation is solved for $G$ given $R$, and means that we
  cannot evolve through an event horizon. We have suppressed the
  function arguments in the table so that it applies for any spatial
  symmetry or none, and any spacetime dimension. ``Gauge evolution
  equations'' denotes the number of equations containing
  $u$-derivatives in the formulation, minus the number of physical
  degrees of freedom (where in the absence of symmetry we count vacuum
  gravity as two degrees of freedom, a scalar field as one, etc).}
\label{table:formulations}
\end{table*}
%%%%%%%%%%%%%%%%%%%%%%%%%%%%%%%%%%%%%%%%%%%%%%%%%%%%%%%%%%%%%%%%%%%%%%%%%%%%%%

We are aware of the following combinations of formulation and gauge
choice used for numerical time evolution in the literature. We hope
that the list of such combinations is complete, but the list of
references is only intended to be indicative:

\begin{itemize}

\item double-null gauge in fe formulation, in spherical
  symmetry: applied to critical collapse
  \cite{HamadeStewart1996}, and to black hole interiors
  \cite{Gnedin93,BurOri97,MurReaTan13};

\item double-null gauge in eR formulation, in
  spherical symmetry: applied to critical collapse
  \cite{Garfinkle1995,ymscalar};

\item Bondi formulation, in spherical symmetry: applied
  to critical collapse \cite{GoldwirthPiran1987,HodPiran97,
GundlachPricePullin1994,PuerrerHusaAichelburg}, and to
  black-hole interiors, \cite{BradySmith95};

\item Bondi formulation, beyond spherical symmetry: applied to axisymmetric weak
  fields \cite{GomezPapadopoulosWinicour1994}, 
  and to gravitational wave extraction
  \cite{ReisswigBishopPollney2013}, see also \cite{Winicour2005};

\item affine formulation, in spherical symmetry: applied to critical
  collapse \cite{CrespoOliveiraWinicour2019}, see also
  \cite{Maedler24};

\item affine formulation, in planar symmetry: applied to
  asymptotically anti-de Sitter spacetimes
  \cite{CheslerYaffe2011};

\item affine formulation, beyond spherical symmetry: 
  applied to asymptotically anti-de Sitter
  spacetimes \cite{CheslerYaffe2014,CheslerLowe2019,Chesler2022}, see
  also \cite{Winicour2013}.

\end{itemize} 

In our own previous work, we have explored the following
additional combinations:

\begin{itemize}

\item ``shifted double null'' gauge in eR formulation, in
  spherical symmetry: applied to critical collapse
  \cite{PortoGundlach2022}, see also \cite{axinull_formulation};

\item ``local shifted Bondi'' gauge in eR formulation, in
  axisymmetry: applied to critical collapse
  \cite{axinull_critscalar}, see also \cite{axinull_formulation}.

\end{itemize}

%%%%%%%%%%%%%%%%%%%%%%%%%%%%%%%%%%%%%%%%%%%%%%%%%%%%%%%%%%%%%%%%%%%%%%%%%%%%%%%

\subsection{Plan of this paper}

%%%%%%%%%%%%%%%%%%%%%%%%%%%%%%%%%%%%%%%%%%%%%%%%%%%%%%%%%%%%%%%%%%%%%%%%%%%%%%%

In the remainder of this paper, we will restrict to spherical symmetry
in 3+1 dimensions, with charged scalar field matter, but the ideas
illustrated here are in principle easily generalised to less symmetry, other
spacetime dimensions, and to vacuum or other matter.

We present an implementation of the five formulations described above
in one code, and we test the three formulations that admit a general
radial gauge in two settings. In the first of these, we evolve on
outgoing null cones that emanate from a regular central worldline. To
fix the radial gauge in the initial data, we either set $R=x$ or $G=1$
at $u=0$. We continue in ``shifted double-null gauge'', {\new see
Eq.~(\ref{sdnshift}) below}, an incremental gauge that is adapted to
critical collapse in spherical symmetry (see also Sec.~III.C of
Paper~I for a definition and \cite{PortoGundlach2022} for a previous
application). As a physics application, we compute the fine-structures
of the mass and charge scaling laws in critical collapse, which have
not been computed before.  In the second setting the left boundary
$x=0$ is not a regular centre but an ingoing null cone, so that the
computational domain is a null rectangle. We continue the initial
gauge choice of $R_{x,}=1$ or $G=1$ as double-null gauge $B=0$.

We do not present any results in the affine and Bondi formulations, as
the affine and Bondi gauges that they require do not allow us to make
the outer boundary future spacelike (in the first setting) or ingoing
null (in the second setting).

In both settings, we present evidence of second-order convergence with
resolution, both within each formulation and between the three pairs
of formulations. With a regular centre, the convergence is not quite
pointwise at the first few grid points, but it is pointwise elsewhere,
and is second-order in any norm. By contrast, on the null rectangle,
where we do not need an expansion, the convergence is cleanly
pointwise.

We devise a simple singularity excision scheme, and present evidence
that it works in the eG and free evolution formulations
(eR cannot evolve through the apparent horizon).

In Sec.~\ref{sec:fieldequations} we give the field equations for the
spherical Einstein-charged-scalar system, in Sec.~\ref{sec:algorithm}
we present our solution algorithm, and in
Sec.~\ref{sec:numericalmethods} our numerical
methods. Sec.~\ref{sec:regularcentre} presents tests on a domain with
a regular centre, including convergence tests, critical collapse, and
singularity excision, and Sec.~\ref{sec:nullrectangle} presents tests
on a null rectangle domain, including convergence tests, collapse, and
singularity excision. We conclude in Sec.~\ref{sec:conclusions}.

%%%%%%%%%%%%%%%%%%%%%%%%%%%%%%%%%%%%%%%%%%%%%%%%%%%%%%%%%%%%%%%%%%%%%%%%%%%%%

\section{The Einstein-Maxwell-charged scalar system}
\label{sec:fieldequations}

%%%%%%%%%%%%%%%%%%%%%%%%%%%%%%%%%%%%%%%%%%%%%%%%%%%%%%%%%%%%%%%%%%%%%%%%%%%%%

\subsection{Field equations in covariant form}

%%%%%%%%%%%%%%%%%%%%%%%%%%%%%%%%%%%%%%%%%%%%%%%%%%%%%%%%%%%%%%%%%%%%%%%%%%%%%

The matter is a complex scalar field $\phi=:\psi+i\chi$ coupled to
electromagnetism. The action, in units where $c=G_N=1$, is
\begin{equation}
S=\int \left({R\over {16\pi}}-{{1\over 2}}D_a\phi
(D^a\phi)^*-{1\over {16\pi}}
F_{ab}F^{ab}\right)\sqrt{-g}\,d^4x,
\end{equation}
where a star denotes the complex conjugate, 
the field strength tensor $F$ in terms of the potential $A$ is
\begin{equation}
F_{ab}:=\nabla_aA_b-\nabla_bA_a,
\end{equation}
and we have introduced the charge-covariant derivative
\begin{equation}
D_a:=\nabla_a+iqA_a.
\end{equation}
Our electromagnetic convention is similar to Gauss units, and our
scalar field convention reduces to the standard one for a real scalar
field $\psi$ if $\chi=0$. These are also the conventions of
\cite{HodPiran97}. The convention of \cite{GellesPretorius25} differs
by the absence of the factor $1/2$ in the scalar field term. The
convention of \cite{MurReaTan13} differs by factors of $16\pi$ in
front of $R$ and $4\pi$ in front of $F_{ab}F^{ab}$.

$\phi$ obeys the wave equation
\begin{equation}
D^aD_a\phi=0.
\end{equation}
The Maxwell equations are
\begin{equation}
\label{divF}
{\nabla_bF^{ab}=4\pi j^a},
\end{equation}
where the charge current is
\begin{eqnarray}
j_a&=&-{iq\over 2}(D_a\phi)^*\phi+c.c. \\
&=&q(\chi\nabla_a\psi-\psi\nabla_a\chi)-q^2A_a(\psi^2+\chi^2).
\end{eqnarray}
It is conserved, $\nabla_a j^a=0$, because $F^{ab}$ is antisymmetric.
The trace-reversed Einstein equations are
\begin{equation}
E_{ab}:=R_{ab}-8\pi S_{ab}=0,
\end{equation}
where the trace-reversed stress-energy tensor is
\begin{eqnarray}
S_{ab}&=& 
{{1\over 2}}\left(D_a\phi(D_b\phi)^*+(D_a\phi)^*D_b\phi\right) \nonumber \\ 
&&+{{1\over 4\pi}}\left(F_{ac}{F_b}^c-{1\over
  4}g_{ab}F_{cd}F^{cd}\right).
\end{eqnarray}

The equations admit the gauge freedom $\phi\to e^{iq\alpha}\phi$,
$A_a\to A_a+\nabla_a \alpha$ for an arbitrary scalar function
$\alpha$, leaving $F_{ab}$ and $D_a \phi$, and hence $j^a$ and
$T_{ab}$, invariant.

%%%%%%%%%%%%%%%%%%%%%%%%%%%%%%%%%%%%%%%%%%%%%%%%%%%%%%%%%%%%%%%%%%%%%%%%%%%%%

\subsection{Field equations in spherical symmetry in null coordinates}

%%%%%%%%%%%%%%%%%%%%%%%%%%%%%%%%%%%%%%%%%%%%%%%%%%%%%%%%%%%%%%%%%%%%%%%%%%%%%

We consider the general metric in spherical symmetry with one null
coordinate $u$,
\begin{equation}
ds^2=-2G\,du(dx+B\,du) +R^2\,d\Omega^2,
\end{equation}
where $G$, $B$ and $R$ are functions of $(u,x)$, and
$d\Omega^2:=d\theta^2+\sin^2\theta\,d\varphi^2$. The surfaces of
constant $u$ are outgoing null cones. The choice of coordinate $x$
along the null cone generators (the radial gauge) is not yet fixed.

Starting from an arbitrary electromagnetic gauge, the particular gauge
\begin{equation} 
A_x(u,x)=0, \qquad A_u(u,0)=0
\end{equation}
can be achieved by the gauge transformation with
\begin{equation}
\alpha(u,x)=-\int_0^u A_u(u',0)\,du'-\int_0^x A_x(u,x')\,dx'.
\end{equation}
From now on, we work exclusively in this gauge, and for conciseness we
rename $A_u$ to $A$. Note that $A_x=0$ can be written geometrically as
$U^aA_a=0$.

We define the ingoing null vector field
\begin{equation}
\label{Xidef}
\Xi:=\partial_u-B\partial_x,
\end{equation}
which is normalised relatively to $U^a$ as
$U^a\Xi_a=-1$. For the scalar
field, we also introduce 
\begin{equation}
\hat\Xi\phi:=(\Xi+iqA)\phi, 
\end{equation}
with real and imaginary parts
\begin{eqnarray}
\label{Xihatpsi}
\hat\Xi\psi&:=&{\rm Re}\,\hat\Xi\phi=\Xi\psi-qA\chi, \\
\label{Xihatchi}
\hat\Xi\chi&:=&{\rm Im}\,\hat\Xi\phi=\Xi\chi+qA\psi.
\end{eqnarray}

The two components of the Maxwell equations can be written in terms of
the local charge function
\begin{equation}
\label{Qdef}
Q:={R^2A_{,x}\over G}.
\end{equation}
The Maxwell equations become
\begin{eqnarray}
\label{Qeqn}
Q_{,x}&=&{4\pi}qR^2(\psi\chi_{,x}-\chi\psi_{,x}), \\
\label{XiQeqn}
\Xi Q&=&{4\pi}qR^2(\chi\hat\Xi\psi-\psi\hat\Xi\chi).
\end{eqnarray}
Hence we have a conserved charge current $\nabla_aj^a=0$ with
\begin{equation}
\label{chargecurrent}
j^u={1\over {4\pi}GR^2}Q_{,x}, \quad j^x=-{1\over {4\pi}GR^2}Q_{,u},
\end{equation}
where an overall constant has been fixed from (\ref{divF}), and
we have used $\sqrt{-g}=GR^2\sin\theta$.

We can write the Einstein and Maxwell equations in terms of $\Xi$,
$\hat\Xi$ and $Q$, without any explicit appearance of $A$, $B$ and
$\partial_u$, with the exception of the Einstein equation
$E_{ux}$. However, we still need to find $B$ in order to recover
$\partial_u$ from $\Xi$, and we need to find $A$ to recover $\Xi$ from
$\hat\Xi$. We find $A$ by integrating (\ref{Qdef}), written as
\begin{equation}
\label{Aeqn}
A_{,x}={GQ\over R^2}.
\end{equation}

In spherical symmetry, the Hawking mass $M$ of a symmetry sphere is a
scalar on the reduced spacetime, and is then also called the
Misner-Sharp mass, or the Kodama mass. It is given by
\begin{equation}
\label{Mdef}
M:={CR\over 2},
\end{equation}
where we define the Hawking compactness $C$ of a symmetry sphere
(equivalent to a point in the reduced spacetime) as
\begin{equation}
\label{Cdef}
C:=1-|\nabla R|^2.
\end{equation}
In our coordinates, $C$ is given by
\begin{equation}
\label{Cexpr}
C(u,x)=1+{2R_{,x}\Xi R\over G}.
\end{equation}

The derivatives of the Misner-Sharp mass are
\begin{eqnarray}
\label{Mxexpr}
M_{,x}&=&-4\pi{R^2 \Xi R\over G}(\psi_{,x}^2+\chi_{,x}^2)+{Q^2 R_{,x}\over 2R^2}, \\
\label{XiMexpr}
\Xi M&=&-4\pi{ R^2 R_{,x}\over G}(\hat\Xi\psi^2+\hat\Xi\chi^2)+{Q^2\Xi
R\over 2R^2}.
\end{eqnarray}
Hence there is a conserved mass curent analogous to
(\ref{chargecurrent}). 

We also define the augmented mass
\begin{equation}
\label{calMdef}
{\cal M}:=M+{Q^2\over 2R}.
\end{equation}
(Other authors call it the renormalised mass.) As $Q$ and $M$ are
scalars on the reduced spacetime, so is ${\cal M}$.  Its derivatives
are
\begin{eqnarray}
\label{calMxexpr}
{\cal M}_{,x}&=&-4\pi{R^2 \Xi R\over G}(\psi_{,x}^2+\chi_{,x}^2)
\nonumber \\ &&
+{4\pi}qQR(\psi\chi_{,x}-\chi\psi_{,x}), 
\\
\Xi {\cal M}&=&-4\pi{ R^2 R_{,x}\over G}(\hat\Xi\psi^2+\hat\Xi\chi^2)
\nonumber \\ &&
+{4\pi}qQR(\chi\hat\Xi\psi-\psi\hat\Xi\chi).
\label{XicalMexpr}
\end{eqnarray}
Hence there is another conserved mass current analogous to
(\ref{chargecurrent}). 

In electrovacuum, defined by $\phi=0$ (in our electromagnetic gauge
choice $A_x=0$), both $Q$ and ${\cal M}$ are constant, and are equal
to the parameters $Q_0$ and ${\cal M}_0$ of the Reissner-Nordstr\"om
(from now on, RN) family of solutions (see Appendix~\ref{appendix:RN}),
while $M$ is not constant unless $Q=0$.  On the other hand, $M$ is
non-decreasing along outgoing null rays, in the sense that $R_{,x}>0$
implies $M_{,x}\ge 0$, both in vacuum and for any matter obeying the
dominant energy condition.

We note that $M$ can be computed from $R$ and $\Xi R$ via (\ref{Mdef})
and (\ref{Cdef}), or by integrating (\ref{Mxexpr}) from $M=0$ at
$R=0$. Similarly, ${\cal M}$ can be computed from $M$ or by
integrating (\ref{calMxexpr}) from ${\cal M}=0$ at $R=0$. We call
these integrated versions of $M$ and ${\cal M}$, $\tilde M$ and
$\tilde{\cal M}$. In the discretised equations, $M$ differs from
$\tilde M$, and ${\cal M}$ from $\tilde{\cal M}$, but by construction
$\tilde M$ retains the non-decreasing property, and $\tilde{\cal M}$
the constant-in-electrovacuum property, even in the presence of
numerical error.

In spherical symmetry, only the components $E_{uu}$, $E_{ux}$,
$E_{xx}$ and $E_+:=E_{\theta\theta}=E_{\varphi\varphi}/\sin^2\theta$
of the trace-reversed Einstein equations are algebraically
independent.  We can write $E_{xx}=0$ as
\begin{equation}
\label{Geqn}
\left(\ln {G\over R_{,x}}\right)_{,x} =4\pi
     {R(\psi_{,x}^2+\chi_{,x}^2)\over R_{,x}},
\end{equation}
which can be integrated for $G$, given $R$, $\psi$ and $\chi$, but
only if $R_{,x}> 0$. Alternatively, we can write it
as
\begin{equation}
\label{Reqn}
R_{,xx}-{G_{,x}\over G}R_{,x} +4\pi(\psi_{,x}^2+\chi_{,x}^2)R=0,
\end{equation}
which can be solved as a second-order linear ODE for $R$, given $G$, $\psi$ and
$\chi$. There is then no restriction on the sign of $R_{,x}$. More geometrically,
$E_{xx}=0$ can be written as
\begin{equation}
\label{UUR}
UUR+4\pi[(U\psi)^2+(U\chi)^2]R=0.
\end{equation}
It is the Raychaudhuri equation for the generators of the coordinate
null cones. 

We can write $E_+=0$ and the real and imaginary parts of the complex wave
equation as
\begin{eqnarray}
\label{XiReqn}
(R\Xi R)_{,x} &=& {-{G\over 2}\left(1-{Q^2\over R^2}\right)}, \\
\label{Xipsieqn}
(R\hat\Xi \psi)_{,x} &=& -(\Xi R)\psi_{,x}-{qQG\chi\over 2R}, \\
\label{Xichieqn}
(R\hat\Xi \chi)_{,x} &=& -(\Xi R)\chi_{,x}+{qQG\psi\over 2R}.
\end{eqnarray}
$E_{ux}=0$ can be written as 
\begin{eqnarray}
\label{calHeqn}
{\cal H}_{,x}&=& 
-{G+2R_{,x}\Xi R\over R^2}+ {{2}GQ^2\over R^4}\nonumber \\ &&
+8\pi (\chi_{,x}\hat\Xi\chi+\psi_{,x}\hat\Xi\psi)
 \\
\label{calHeqnbis}
&=&{-{2G{\cal M}\over R^3}+{3GQ^2\over R^4}
+8\pi(...)},
\end{eqnarray}
where 
\begin{equation}
\label{calHdef}
{\cal H}:=B_{,x}-\Xi\ln G
\end{equation}
is the only place where $B$ appears in the Einstein equations, other
than in the combination $\Xi$. $E_{ux}$ is a combination of
derivatives of the other Einstein equations, and is therefore
sometimes called the ``redundant'' equation, but in the eG and fe
formulations we use it to obtain $\Xi\ln G$, given an arbitrary $B$,
while in the affine formulation we use it to obtain $B$, given $G=1$.

The case where $x=0$ is a regular centre $R=0$ is considered in
Appendix~\ref{appendix:regularcentre}. From regularity, $M$, $Q$,
${\cal M}$, $R\Xi R$, $R\Xi\psi$ and $R\Xi\chi$ all vanish at $x=0$,
$A$ vanishes there as an electromagnetic gauge choice, and ${\cal H}$
vanishes because of regularity and because we choose $u$ to be proper
time at the centre. Hence we have trivial boundary conditions for the
integration of all hierarchy equations.

The remaining Einstein equation $E_{uu}=0$ can be written as
\begin{equation}
\label{EEuu}
\Xi\Xi R+{\cal H}\Xi R+4\pi
R(\hat\Xi\psi^2+\hat\Xi\chi^2)=0.
\end{equation}
It is the Raychaudhuri equation on the ingoing null geodesics.  We do
not use it in our time evolution scheme with a regular centre, but it
acts as a on constraint on the data we can impose on a left boundary
that is ingoing null. In analogy to (\ref{UUR}), we can write it more
geometrically as
\begin{equation}
\label{XXR}
X\!XR+4\pi[(\hat X\psi)^2+(\hat X\chi)^2]R=0,
\end{equation}
where 
\begin{equation}
X:=\bar G^{-1}\Xi, \qquad \hat X:=\bar G^{-1}\hat\Xi.
\end{equation}
$X$ is tangent to the affinely parameterised ingoing null geodesics (as $U$
is to the outgoing ones), and $\bar G$ is defined by
\begin{equation}
\Xi\ln \bar G=-{\cal H},
\end{equation}
{or equivalently by
\begin{equation}
\Xi\ln \left({\bar G\over G}\right)=-B_{,x}.
\end{equation}
In double-null gauge $B=0$, we have $\bar G=G$ and
$X=G^{-1}\partial_u$, in analogy to $U=G^{-1}\partial_x$ (the latter
holds in any radial gauge). 

In initial data on an outgoing null cone with regular centre, we
always choose $\Xi R<0$ and hence $XR<0$ on $u=0$. From
(\ref{XXR}) we see that $XXR\le 0$, so $XR<0$ and hence $\Xi R<0$ on
the entire domain of dependendence of these initial data. This is not
obvious from (\ref{XiReqn}), whose right-hand side becomes negative
when $R<|Q|$, and so, as we solve (\ref{XiReqn}) but not (\ref{XXR}),
$\Xi R<0$ may be violated by numerical error.}

As $R\ge 0$, $G>0$  and $\Xi R<0$, $\rho_+$ given above in
(\ref{defrhoplus}) has the same sign as $R_{,x}$. Hence a symmetry
sphere is outer-trapped if and only if $R_{,x}<0$, or
equivalently $C>1$.

The choice of coordinate $x$ for $u>0$ is controlled incrementally by
the choice of $B$. This plays a role similar to the radial shift
vector component in the ADM formulation. To see this, note that if we make a
transformation from double-null coordinates $(u,v)$ to
general null coordinates $(u,x)$, parameterised by $v=v(u,x)$, then
\begin{equation}
-2\bar G\,du\,dv =-2G\,du(dx+B\,du)
\end{equation}
holds if and only if
\begin{eqnarray}
G&=&\bar G v_{,x}, \\
\label{vux}
v_{,u}&=&B v_{,x},
\end{eqnarray}
regardless of spacetime dimension or symmetry, and so $v$ is advected
in $x$ with speed $-B$.

Our key diagnostics are the Hawking compactness $C$, Hawking mass $M$,
charge $Q$ and augmented mass ${\cal M}$ defined above. Extending the
definition in Paper~II, we define
\begin{equation}
T:=\max_\text{whole spacetime}|(D_a\phi)^*D^a\phi|
\end{equation}
as an additional diagnostic.  Note that $\max|{R_a}^a|=8\pi T$. This
gives us a curvature quantity that scales on the dispersion side of
the collapse threshold.

%%%%%%%%%%%%%%%%%%%%%%%%%%%%%%%%%%%%%%%%%%%%%%%%%%%%%%%%%%%%%%%%%%%%%%%%%%%%%

\section{Algorithm}
\label{sec:algorithm}

%%%%%%%%%%%%%%%%%%%%%%%%%%%%%%%%%%%%%%%%%%%%%%%%%%%%%%%%%%%%%%%%%%%%%%%%%%%%%

%%%%%%%%%%%%%%%%%%%%%%%%%%%%%%%%%%%%%%%%%%%%%%%%%%%%%%%%%%%%%%%%%%%%%%%%%%%%%

\subsection{Initial data and gauge choice}

%%%%%%%%%%%%%%%%%%%%%%%%%%%%%%%%%%%%%%%%%%%%%%%%%%%%%%%%%%%%%%%%%%%%%%%%%%%%%

To set initial data, we specify $\psi(x)$ and $\chi(x)$ freely. We
then either initialise $G=1$ and solve (\ref{Reqn}) to initialise $R$,
or we initialise $R=x$ and solve (\ref{Geqn}) to initialise $G$. These
are physically inequivalent data for the same functions $\psi(x)$ and
$\chi(x)$, as they correspond to different geometrically defined data
functions $\psi(R)$ and $\chi(R)$, or equivalently $\psi(\lambda)$ and
$\chi(\lambda)$.

In Bondi gauge $R=x$, the identity $R_{,u}=0$ determines $B$ as
\begin{equation}
\label{Bondishift}
B_\text{Bondi}=-{\Xi R\over R_{,x}}.
\end{equation}
All our gauge choices keep $R=0$ at $x=0$, and so $R_{,u}=0$ at the
centre. With $u$ defined to be proper time there, we have
\begin{equation}
\label{centralshift}
B(u,0)=B_\text{Bondi}(u,0)={1\over 2R_{,x}(u,0)},
\end{equation}
see also Eq.~(\ref{BBC}) below. 

In affine gauge $G=1$, the identity $\Xi G=0$ determines $B$ as
\begin{equation}
\label{affineshift}
B_\text{affine}=B_\text{Bondi}(u,0)+\int_0^x{\cal H}\,dx', 
\end{equation}
where we have used (\ref{centralshift}). 

For critical collapse in spherical symmetry, we use a radial gauge
where $B$ is a linear function of $x$, fixed to coincide with
Bondi gauge at $x=0$ (so that $R=0$ remains at $x=0$) and to vanish at
$x=x_0$ (so that $x=x_0$ is an ingoing null cone). This gives
\begin{equation}
\label{sdnshift}
B_\text{sdn}(u,x)=B_\text{Bondi}(u,0)
\left(1-{x\over x_0}\right).
\end{equation}
It is easy to see from (\ref{vux}) that if $v=x$ on $u=0$ and $B$ is
proportional to $x$, then $v$ remains linear in $x$, with $v=x_0$ at
$x=x_0$ and $R=0$ at $x=0$, for all $u\ge 0$. This simple linear
relation between $v$ and $x$ motivates the name ``shifted
double-null'' (sdn) gauge.

%%%%%%%%%%%%%%%%%%%%%%%%%%%%%%%%%%%%%%%%%%%%%%%%%%%%%%%%%%%%%%%%%%%%%%%%%%%%%

\subsection{Boundary conditions at $x=0$}

%%%%%%%%%%%%%%%%%%%%%%%%%%%%%%%%%%%%%%%%%%%%%%%%%%%%%%%%%%%%%%%%%%%%%%%%%%%%%

When $x=0$ is the regular central worldline $R=0$, $x=0$ is a
coordinate singularity, and we cannot impose simple boundary
conditions there. To start up the integration of the hiercharchy
equations, we expand in powers of $x$ instead.

At a given $u$, we fit the evolved quantities $\psi$, $\chi$, $R/x$
and/or $G$ either to linear functions of $x$, such as $\psi\simeq
\psi_{(0)}+\psi_{(1)}x$, or to quadratic ones, such as $\psi\simeq
\psi_{(0)}+\psi_{(1)}x+\psi_{(2)}x^2$. (The notation here is
compatible with our notation $\psi_{l(n)}$ in Papers~I and II, where $l$
denotes the spherical harmonic and $n$ the power of $x$.)

$R_{(1)}$, $G_{(0)}$ and ${\cal H}_{(0)}$ are not determined by the
equations. However, we show in Appendix~\ref{appendix:regularcentre}
that at a regular centre, and with $u$ proper time at the centre,
they obey $G_{(0)}=R_{(1)}$, $B_{(0)}=1/(2R_{(1)})$ and ${\cal
  H}_{(0)}=0$.

The full expressions for the expansion coefficients are given
in Appendix~\ref{appendix:expansions}.

On the other hand, if $x=0$ is a timelike or null boundary, we
can specify boundary conditions for the hierarchy equations there,
subject to constraints. For the case where $x=0$ is an ingoing null
cone, we discuss this in detail in Appendix~\ref{appendix:ingoingdata}.

%%%%%%%%%%%%%%%%%%%%%%%%%%%%%%%%%%%%%%%%%%%%%%%%%%%%%%%%%%%%%%%%%%%%%%%%%%%%%

\subsection{Time evolution algorithm}
\label{section:algorithm}

%%%%%%%%%%%%%%%%%%%%%%%%%%%%%%%%%%%%%%%%%%%%%%%%%%%%%%%%%%%%%%%%%%%%%%%%%%%%%

We use the ``method of lines'' framework, where one discretises
separately in space and time. We break up each sub-timestep into the
following steps, where Steps 1 and 3 depend on the formulation.  For
completeness, we include the affine and Bondi formulations in the
following pseudo-code, even though we do not use them for any
numerical tests in this paper.

{\bf Step 0}: Assume that on a null cone of constant $u$, the
functions $\psi(x)$, $\chi(x)$, and $G(x)$ and/or $R(x)$ are
given. When $G$ and $R$ are both given, they must obey the constraint
(\ref{Reqn}).  With a regular centre, we fit these functions to
power-series expansions in $x$ about $x=0$.

{\bf Step 1a}: In the eR or Bondi formulation, we solve the
Raychaudhuri equation in its form (\ref{Geqn}) for $G$ by
integration, and set the fitting coefficients of $G$ consistently. 

{\bf Step 1b}: Alternatively, in the eG or affine formulation,
we solve the Raychaudhuri equation in its form (\ref{Reqn}) as a
homogeneous linear second-order ODE for $R$, and we set the fitting
coefficients of $R$ consistently. 

{\bf Step 1c}: Alternatively, in the fe formulation we do nothing. From
numerical error, both in the evolution of $R$ and $G$, and in the
fitting, the fitting coefficients of $R$ and $G$ will not be exactly
compatible with the Raychaudhuri equation, so we need to make an
arbitrary choice which set of fitting coefficients we choose for
expanding the solution of the hierarchy equations at the centre. {
  The choice made in Appendix~\ref{appendix:expansions} works.}

{\bf Step 2}: We now solve (\ref{Qeqn}) for $Q$ and then (\ref{Aeqn})
for $A$ by integration. Next we solve (\ref{XiReqn}) for $\Xi R$ by
integration, followed by (\ref{Xipsieqn}) and (\ref{Xichieqn}) for
$\hat\Xi\psi$ and $\hat\Xi\chi$, and hence $\Xi\psi$ and
$\Xi\chi$. Finally, we solve (\ref{calMxexpr}) for ${\cal M}$ (which,
obtained thus, we call $\tilde {\cal M}$) and then (\ref{calHeqnbis})
for ${\cal H}$. (In the eR and Bondi formulations, ${\cal H}$ is not
required, and we compute it only as a diagnostic.)

{\bf Step 3a}: In the affine formulation, we compute $B$ from
(\ref{affineshift}).

{\bf Step 3b}: Alternatively, in the Bondi formulation, $B$ is given
explicitly by (\ref{Bondishift}).

{\bf Step 3c}: Alternatively, in the eR, eG and fe formulations we set $B$
freely. {For critical collapse in spherical symmetry, we use
sdn gauge (\ref{sdnshift}). On a null rectangle we use double-null
gauge $B=0$.}

{\bf Step 4}: We find the $u$-derivatives of $\psi$, $\chi$, $R$
and/or $\ln G$ from their $\Xi$-derivatives and $x$-derivatives, using
(\ref{Xidef}). The $x$-derivatives are upwinded depending on the sign
of $B$. At $x=0$ this makes the upwind derivative a right derivative,
and at $x_\text{max}>x_0$ a left derivative.

{\bf Step 5}: Using an ODE integrator at each grid point (the
method of lines), we
evolve $\psi$, $\chi$, and one or both of $R$ and $\ln G$, from $u$ to
$u+\Delta u$. This returns us to Step 0.

%%%%%%%%%%%%%%%%%%%%%%%%%%%%%%%%%%%%%%%%%%%%%%%%%%%%%%%%%%%%%%%%%%%%%%%%%%%%%

\section{Numerical methods}
\label{sec:numericalmethods}

%%%%%%%%%%%%%%%%%%%%%%%%%%%%%%%%%%%%%%%%%%%%%%%%%%%%%%%%%%%%%%%%%%%%%%%%%%%%%

\subsection{Discretization}
\label{section:discretization}

%%%%%%%%%%%%%%%%%%%%%%%%%%%%%%%%%%%%%%%%%%%%%%%%%%%%%%%%%%%%%%%%%%%%%%%%%%%%%

With $x=R=0$ a regular centre, we use an equally spaced centered grid
$x_i=i\Delta x$, $i=1...N_x$, so that $x=0$ is not on the grid. With
$x=0$ an ingoing null cone, we use $x_i=i\Delta x$, $i=0,1...N_x$.

The ODEs in $x$ can be solved by integration using an integrating
factor. They can therefore be written as $g_{,x}=F(f,f_{,x})$, where the $f$ are
already known and $g$ is to be determined. We use the explicit
integration scheme
\begin{equation}
\label{twopointstencil1}
g_i\simeq g_{i-1}+F\left(f_{i-{1\over 2}},(f_{,x})_{i-{1\over 2}}\right)\,\Delta x ,
\end{equation}
with
\begin{eqnarray}
\label{twopointstencil2}
f_{i-{1\over 2}}&\simeq& {f_i+f_{i-1}\over 2}, \\
\label{twopointstencil3}
(f_{,x})_{i-{1\over 2}}&\simeq& {f_i-f_{i-1}\over \Delta x}.
\end{eqnarray}
This scheme is explicit, is second-order accurate by construction, and is
causal in the sense that the vector of variables $g_i$ depends only on
$g_j$ with $j<i$ and $f_j$ with $j\le i$: information propagates along
a null cone of constant $u$ only towards increasing $x$.

To solve the Raychaudhuri equation (\ref{Reqn}) for $R$, we write it
in the first-order form
\begin{eqnarray}
\label{Rxeqn}
R_{,x}&=&G\,V,\\
\label{Vxeqn}
V_{,x}&=&-{K \over G}\,R, 
\end{eqnarray}
with
\begin{eqnarray}
V&:=&{R_{,x}\over G}, \\
K&:=&4\pi(\psi_{,x}^2+\chi_{,x}^2). 
\end{eqnarray}
When we solve (\ref{Vxeqn},\ref{Rxeqn}) for $(V,R)$, the
coefficients $G$ and $K$ are already known.

$K$ is naturally given at midpoints, but $G$ needs to be interpolated
there using (\ref{twopointstencil2}). We discretize this system, using
a 2-point stencil, to second-order accuracy as
\begin{eqnarray}
{R_i-R_{i-1}\over\Delta x}&=&G_{i-{1\over 2}}{V_i+V_{i-1}\over 2}, \\ 
{V_i-V_{i-1}\over\Delta x}&=&-{K_{i-{1\over 2}}\over G_{i-{1\over 2}}}
{R_i+R_{i-1}\over 2}.
\end{eqnarray}
This can be solved for $R_i$ and
$V_i$, giving us the explicit scheme
\begin{equation}
\left(\begin{array}{c}R_i \\ V_i \\
\end{array}\right)
={1\over 1+{{\cal K}}}\left(\begin{array}{cc} 1-{{\cal K}} & 
{\cal G}  \\
-{4{{\cal K}}\over {\cal G}} & 
1-{{\cal K}} \\
\end{array}\right)
\left(\begin{array}{c}R_{i-1} \\ V_{i-1} \\
\end{array}\right),
\end{equation}
where 
\begin{eqnarray}
{\cal K}&:=&K_{i-{1\over 2}}{\Delta x^2\over 4}, \\
{\cal G}&:=& G_{i-{1\over 2}}\Delta x .
\end{eqnarray}
The two complex eigenvalues of this linear map have absolute value 1
for any ${{\cal K}}\ge 0$, so that the method is unconditionally
stable. The method is exact for $K=0$. It has no amplitude error when
$K_{,x}=G_{,x}=0$, in the sense that then $(KR^2+G^2V^2)_{,x}=0$ in the
discretisation, as it is in the continuum.

$V(x)$ changes sign at most once (from positive to negative) over the
$x$-domain. This means that $K\lesssim x_\text{max}^{-2}$, and so
${{\cal K}}\lesssim N_x^{-2}\ll 1$. Hence we can expect the numerical
error to be reasonably small.

For comparison, we also use the second or fourth-order Adams-Bashforth
method. These need the vector of right-hand sides $F_i$ at grid
points. We discretize $x$-derivatives on grid points as
\begin{equation}
(f_{,x})_i\simeq {f_{i+1}-f_{i-1}\over 2\Delta x}
\end{equation}
where this is defined, and the standard one-sided three-point stencils
at the left and right boundaries $i=1$ and $i=N_x$.  This is
also second-order accurate but not causal.

When we solve (\ref{Geqn}) for $G$, we first find $-\ln V$ on grid
points by integration, where $V:=R_{,x}/G$. (In Papers~I-III, we used
the variables $g:=1/V$ and $\gamma:=\ln g$. We have replaced them with
$V$ as they are defined only for $R_{,x}>0$.) We then find $V$ at grid
points and hence at mid-points, and hence $G=R_{,x}/V$ at mid-points
in a causal and second-order accurate way, given that $R_{,x}$ is
known to second order and causally at mid-points.

We linearly interpolate $G$ from the mid-points to grid points (and we
linearly extrapolate to the grid points $i=1$ and $i=N_x$). This is
second-order accurate but with an additional error from the
interpolation, and it is no longer causal. However, we need $G$ at
grid points only for output and to initialise $\ln G$ from initial
data in Bondi gauge. The non-causality does not matter for output and
initialisation. To evaluate the compactness $C$ when $G$ is
constrained, we use $V$ on grid points. 

$G$ varies rapidly in $u$ and $x$, but is constrained to be strictly
positive. To avoid it becoming negative through numerical error, we
evolve $\ln G$ rather than $G$. We find $G$ at grid points by
exponentiating, and hence on mid-points by linear interpolation. To
evaluate the compactness $C$ when $G$ is evolved, we us $R_{,x}$ on
gridpoints, which is not causal. Again we do not see an adverse
effect.

In the evolution equations 
\begin{equation}
\label{transport}
f_{,u}=\Xi f + B f_{,x}
\end{equation}
where $f$ stands for $\psi$, $\chi$, $R$ and $\ln G$, we discretize
the transport term $f_{,x}$ by second-order upwinding, depending on
the sign of $B$. If our solution method for the hierarchy equations is
also causal, as defined above, our complete numerical method is
strictly causal, in the sense that all grid points are computed only
from grid points in their causal past. In particular, with $B<0$ at
$i=N_x$ and $i=N_x-1$ , the outer boundary is treated numerically
like any other point.

In any gauge where $R=x$ stays at $x=0$ we need to evaluate $R_{,x}$
at the centre to impose (\ref{centralshift}). Given that the upwinded
transport term $BR_{,x}$ should cancel $\Xi R$ to give us $R_{,u}=0$
at the centre, we compute $R_{,x}$ at the centre as the second-order
upwind derivative
\begin{equation}
R_{(1)}:=R_{,x}(u,0)\simeq {2 R_1-{1\over 2}R_2\over \Delta x}.
\end{equation}
Other choices result in an instability at the centre.

We solve the $f_{,u}=F[f]$ evolution equations using a Runge-Kutta ODE
solver (typically second-order) in $u$, and we solve the ODEs in $x$
that give $F[f]$ at each sub-timestep. 

%%%%%%%%%%%%%%%%%%%%%%%%%%%%%%%%%%%%%%%%%%%%%%%%%%%%%%%%%%%%%%%%%%%%%%%%%%%%%

\subsection{Quantities that scale as a power of $x$ at the centre}

%%%%%%%%%%%%%%%%%%%%%%%%%%%%%%%%%%%%%%%%%%%%%%%%%%%%%%%%%%%%%%%%%%%%%%%%%%%%%

The quantity $Q$ is obtained by integration of (\ref{Qeqn}). From its
definition, the quantity $Q/R^3$, representing an average charge
density, is expected to be smooth, and to be non-zero at the
centre. Numerically, this smoothness is not guaranteed. We can improve
regularity by discretizing
\begin{equation}
\label{myfac}
\int_{x_{i-1}}^{x_i}f\,R^2\,dx\simeq {1-\epsilon+{\epsilon^2\over 3}\over
  1-\epsilon+{\epsilon^2\over 4}} \ f_{i-1/2} \ (R_{i-1/2})^2 \ \Delta x,
\end{equation}
where $\epsilon:=1/i$. {This is} exact when $R/x$ and $f$ are
constant. Applying this to the computation of $Q$ does not make a big
difference for $Q$ itself, but it does for $A$, which is the integral
of $Q/R^2$. We apply the same correction factor in the integral for
$\tilde M$, to correct for a naive midpoint value of $Q/R\sim x^2$.

For a similar reason, whereever we need to evaluate $M/R^3$, $Q/R^2$ and similar
quantities, in the mid-point rule we take the average of 
$M/R^3$ at $x_i$ and $x_{i-1}$, rather than averaging $M$ and $R$. 

Finally, a more accurate way of evaluating (\ref{calHeqn}) is to
rewrite it as (\ref{calHeqnbis}), and then evaluate ${\cal M}$ 
  in the first term as $\tilde {\cal M}$, the integral expression for
${\cal M}$. We find heuristically that this is essential for ${\cal
  H}$ and hence $\Xi\ln G$ to converge with resolution.

%%%%%%%%%%%%%%%%%%%%%%%%%%%%%%%%%%%%%%%%%%%%%%%%%%%%%%%%%%%%%%%%%%%%%%%%%%%%%

\subsection{Choice of time step}

%%%%%%%%%%%%%%%%%%%%%%%%%%%%%%%%%%%%%%%%%%%%%%%%%%%%%%%%%%%%%%%%%%%%%%%%%%%%%

In the spherical and axisymmetric scalar critical collapse simulations
of Paper~II we used the timestep $\Delta u=\Delta u_1$, where
\begin{equation}
\label{doublenull}
\Delta u_1 := C_1 \min_x\left|{R_{,x}\over \Xi R}\right|\, \Delta x.
\end{equation}
(In axisymmetry, the minimum was taken also over the angular
coordinate $y$). As is clear from the formula, this restricts the
change of $R$ with $\Delta u$ in the ingoing null direction, compared
to the change with $\Delta x$ in the outgoing null direction. $\Delta
u=\Delta u_1$ with $C_1=0.1$ was used in all simulations in Paper~II
(although this was not stated there). These simulations used the eR
formulation, and were stopped as soon as $\max_x C\ge 0.99$. As $C=1$
implies $R_{,x}=0$, $\Delta u_1\to 0$ as $\max_x C\to 1_-$.

The hierarchy equations are in effect ODEs in $x$, while the time
update takes the form (\ref{transport}). Hence a necessary condition
for the stability of our explicit scheme is that these advection
equations obey the Courant-Friedrichs-Levy (CFL) condition. We
guarantee this by restricting the timestep to
\begin{equation}
\label{shiftonly}
\Delta u_2 := C_2 \left(\max_x\left|B\right|\right)^{-1}\, \Delta x
\end{equation}
for some $C_2\lesssim 1$. In all our gauges $R_{,u}=0$ at $R=0$, and
so (\ref{doublenull}) and (\ref{shiftonly}) with $C_1=C_2$ are
equivalent at the centre.

Since Paper~II, we have realised that using $\Delta u=\Delta u_1$ may
be an unnecessarily small timestep. We have found that $\Delta
u=\Delta u_2$ with $C_2=0.1$ also leads to stable evolutions up to
close to the apparent horizon, but with fewer timesteps overall.

As the apparent horizon is approached, $\Delta u_1\to 0$, and so this
timestep does not allow us to evolve through the horizon. The
alternative timestep $\Delta u_2$ remains finite, and the evolution
remains stable, but as the solution changes very rapidly with $u$ at
the horizon, $\Delta u_2$ gives large errors there. (In
Appendix~\ref{appendix:geodesicpeeling} we remind the reader that $R$
on constant-$u$ time slices deviates from the horizon radius
exponentially in $x$, and this in turn means that $\Xi R$ near the
horizon grows exponentially in $x$.)

This problem was already noted by Burko and Ori in \cite{BurOri97} in
simulations in spherical symmetry in double null coordinates. They
were able to address it in an elegant way: their code advances an
ingoing null slice in $v$, so that, for computational purposes, $v$ is
the ``time'' coordinate, not $u$. The necessary mesh refinement in $u$
near the horizon is then made adaptively, by splitting grid cells in
$u$ near $u=u_h$, increasingly finely as $v$ increases.

In our framework, we could in principle introduce adaptive mesh
refinement to better resolve the event horizon, but this goes against
our main motivation for using null coordinates in the first place,
which is to resolve type-II critical collapse down to arbitrarily small scales
by adapting the coordinates to self-similarity, without the need for
explicit adaptive mesh refinement. 

Hence to maintain accuracy through $u=h_h$, we just reduce the global
timestep near the horizon. This works because near the horizon the
solution changes rapidly in $u$ but not in $x$. In type-II critical
collapse, most of the computation is in the approximately self-similar
phase, and so is not affected by this relatively inefficient method of
mesh refinement. After some experimentation we have come up with the
timestep
\begin{equation}
\label{shiftandRtres}
\Delta u_4=\min(\Delta u_2, \Delta u_3),
\end{equation}
where
\begin{equation}
\Delta u_3 := C_3 {\max_x R \over \max_x |R_{,u}|},
\end{equation}
with $C_3=0.001$.  Note $\Delta u_3$ and hence $\Delta u_4$ is not
proportional to $\Delta x$, in contrast to $\Delta u_1$ and $\Delta
u_2$.

If we want to evolve through a horizon, where $R_{,x}=0$, we cannot
use the numerical time step $\Delta u_1$. In double-null gauge $B=0$,
the time step $\Delta u_2$ is not defined. The time steps $\Delta u_3$
and $\Delta u_4$ cannot be used for convergence testing as $\Delta
u_3$ is independent of $\Delta x$. As an alternative to all these, for
example if we want to test convergence of a solution in
  double-null gauge where $R_{,x}=0$ occurs, we also use the
simple fixed timestep
\begin{equation}
\Delta u_0:=C_0\Delta x.
\end{equation}

%%%%%%%%%%%%%%%%%%%%%%%%%%%%%%%%%%%%%%%%%%%%%%%%%%%%%%%%%%%%%%%%%%%%%%%%%%%%%

\subsection{Singularity excision}

%%%%%%%%%%%%%%%%%%%%%%%%%%%%%%%%%%%%%%%%%%%%%%%%%%%%%%%%%%%%%%%%%%%%%%%%%%%%%

{\new 
In the integration of the  hierarchy equations from $x=0$, information
propagates only from  smaller to larger $x$. The same  is true for the
time  evolution  equations  (\ref{transport}),  whereever  $B<0$.  Our
discretisation of the hierarchy  and evolution equations respects this
causality.} In spherical  symmetry, we can therefore use  a crude type
of singularity excision where we excise gridpoints $i\ge i_\text{exc}$,
where $i_\text{exc}$ is  the first grid point where  either $R_i<0$ or
$R_i<R_{i-1}/2$.  (The  second criterion  makes  sure  that the  first
non-excised point is not too close  to $R=0$.) It is essential that we
re-determine   $i_\text{exc}$  after   each   time   we  have   solved
(\ref{Reqn})  for  $R$,  that  is,  in  all  sub-timesteps.  Obviously
$i_\text{exc}$ is not allowed to decrease from one sub-timestep to the
next, or from one full timestep to the next.

Once excision starts, the excision radius
$x_\text{exc}(u):=i_\text{exc}(u)\Delta x$ decreases quickly to
zero. Once it is smaller than $x_0$, our upwind differencing is still
to the right at the outer boundary. To avoid this, we adjust $x_0$ so
that it remains the same fraction of the excision radius
$x_\text{exc}$ as it was originally to $x_\text{max}$.

%%%%%%%%%%%%%%%%%%%%%%%%%%%%%%%%%%%%%%%%%%%%%%%%%%%%%%%%%%%%%%%%%%%%%%%%%%%%%

\section{Tests with a regular centre}
\label{sec:regularcentre}

%%%%%%%%%%%%%%%%%%%%%%%%%%%%%%%%%%%%%%%%%%%%%%%%%%%%%%%%%%%%%%%%%%%%%%%%%%%%%

\subsection{Charged initial data}

%%%%%%%%%%%%%%%%%%%%%%%%%%%%%%%%%%%%%%%%%%%%%%%%%%%%%%%%%%%%%%%%%%%%%%%%%%%%%

We set initial data of the form
\begin{equation}
\label{chargedinitialdata}
\psi(0,x)+i\chi(0,x)
=p {\cal A}\,e^{-\left({x-x_c\over d}\right)^2+i\omega(x-x_c)}, 
\end{equation}
so that for $\omega=0$ the scalar field is purely real, and hence for
small $\omega$ we expect the charge to be proportional to
$\omega$. With a charged scalar field, $q$ sets an inverse length
scale in the field equations, so $qx_c$, $qd$ and $q/\omega$ are all
relevant dimensionless parameters. This also means that changing $q\to
\lambda q$ is equivalent to $x_c\to x_c/\lambda$, $d\to d/\lambda$,
and $\omega\to\lambda\omega$, and similarly for any other length
scales in the initial data. Therefore, it may be convenient to
  fix a length scale in the initial data, such as one of $x_c$, $d$
  and $\omega$ in (\ref{chargedinitialdata}), and consider $q$ as a
  parameter of the initial data in its place (rather than a constant
  of nature). Thus, in one test below, we keep $x_c$, $d$ and $\omega$
  the same and vary $q$. Varying $q$ has the conceptual advantage that
  for given initial data $\psi(x)$, $\chi(x)$ and $R(x)$ or $G(x)$,
  $Q(x)$ in the initial data is manifestly proportional to $q$.

%%%%%%%%%%%%%%%%%%%%%%%%%%%%%%%%%%%%%%%%%%%%%%%%%%%%%%%%%%%%%%%%%%%%%%%%%%%%%

\subsection{Convergence tests}

%%%%%%%%%%%%%%%%%%%%%%%%%%%%%%%%%%%%%%%%%%%%%%%%%%%%%%%%%%%%%%%%%%%%%%%%%%%%%

For convergence testing, we have focused on the initial data
(\ref{chargedinitialdata}) with $x_c=0.5$, $d=0.5$, $\omega=2.0$, and
the coupling constant $q=1/\sqrt{4\pi}$. We complete the initial
data either with $G=1$, and then set $p{\cal A}=0.15$, or with $R=x$,
and then set $p{\cal A}=0.2$. These amplitudes are chosen so that
$\max_xC\simeq 0.5$ in the initial data at $x\simeq 1.2$.  

Our numerical domain is determined by the choices $x_\text{max}=4.0$
and $x_0=3$. Both initial data sets collapse promptly. In the
evolution $\max_xC\simeq 0.9$ is reached at $u\simeq 0.6$ at $x\simeq
1.2$, and the event horizon is crossed soon after. In these two
solutions, the time step does not need to be reduced very much to
resolve the horizon (in contrast to near-critical collapse). For each
set of initial data, we evolve with the eR, eG and fe formulations. We
thus have six combinations of initial data and formulation to
consider.

We evolve with $N_x=100\cdot 2^n$ grid points for $n=0...8$. {We
  choose the timestep $\Delta u_4$ with parameters $C_2=0.1$,
  $C_3=10^{-3}$.} We output all variables at intervals $\Delta u=0.1$,
adjusting the last timestep before each output time down in order to
hit it exactly. We evolve up to horizon formation, and compare
resolutions up to the last output time before that. Our convergence
testing methods are summarised in
Appendix~\ref{appendix:convergencetesting}. Near the horizon, the
timestep $\Delta u_4$ can be smaller than the timestep $\Delta u_2$
that is by definition proportional to $\Delta x$. {This does not seem
to affect second-order convergence, presumably because the finite-differencing
error in $x$ dominates over that in $u$.}

We have used convergence testing to experiment with fitting expansions
near $x=0$ of the evolved functions $\psi$, $\chi$, $R/x$ and/or $G$
to different numbers of points, and to linear or quadratic functions,
and independently with expanding the solutions of the hierarchy
equations near $x=0$ to first or second order, and for different
numbers of points before our midpoint integration rule is applied.
For this experimentation, we have found it helpful to first focus on
the initial data, solving only the hierarchy equations, and to make
decisions for one hierarchy equation at a time, in the order in which
they are solved. 

We eventually decided to expand some variables to second order, in the
sense that we expand as far as we can using quadratic fits of $\psi$,
$\chi$, $R/x$ and/or $G$. We then use the results of those quadratic
fits, even where we expand only to first order. With that settled, we
experimented with the order of expansion and the number of points to
expand to, for each variable, in the order in which they are
computed. We found that for all variables, expanding only to $i=1$,
and finding $i=2,3...$ by integration, was optimal. For most
variables, expanding only to first order was optimal. These are the
variables for which the errors $h^2f_2$ and $h^3f_3$ defined in
(\ref{Richardson}) below are finite
at $x=0$.

When expanding $R\hat\Xi\psi$ and $R\hat\Xi\chi$ to first order, we
found that the errors $f_2$ and $f_3$ of $\hat\Xi\psi$ and
$\hat\Xi\chi$ diverge as $x^{-1}$. This is not in itself a problem, as
$h^2f_2(x_i)\simeq h^2/(ih)=h/i$, so even at $i=1$ the error goes down
as $h$. However, we find that by expanding $R\hat\Xi\psi$ and
$R\hat\Xi\chi$ to second order, we can essentially remove the $h^2f_2$
error, so that the error is dominated by $h^3f_3$. This reduces the
actual error at $i=1$ by two or three orders of magnitude already at
$h=0.04$, and by more at higher resolution.

We have also output the quantities $Q/R^3$, $M/R^3$ and $\tilde
M/R^3$ and tested their convergence. Again, they converge to second
order, but the errors go as $x^{-1}$, $x^{-2}$ and $x^{-1}$,
respectively, at the centre. This is why we use $\tilde M/R^3$ where
we need to evaluate $M/R^3$. The correction factors in integration $Q$
and $\tilde M$ in Eq.~(\ref{myfac}) are essential for reducing the
error in $Q/R^3$ and $\tilde M/R^3$ from $x^{-2}$ to $x^{-1}$.

Comparing $M$ computed from (\ref{Mdef}) and (\ref{Cdef}) with $\tilde
M=\int M_{,x}\,dx$, where $M_{,x}$ is given by (\ref{Mxexpr}), or
${\cal M}$ with $\tilde{\cal M}$, tests the consistency of several
of our field equations, as well as their discretizations.

With the methods discussed above, we achieve, roughly speaking,
second-order pointwise convergence almost until the horizon is
crossed. (We lose convergence at the horizon simply because some variables
change very rapidly across the horizon, while the $u$-value of the horizon
depends on the resolution.) 

To state our convergence results more precisely, it is useful to split
all variables into {two} groups.  The variables in the first group,
including $R$, $Q$, $A$, $M$, $\tilde M$, and $V$ (in eR and eG) show
clear pointwise second-order convergence including at the centre.

The variables in the second group, including $\psi$, $\chi$, $\ln G$,
$\ln(-2\Xi R)$, $\Xi\psi$, $\Xi\chi$, $C$, $x^2M/R^3$,
$x\tilde M/R^3$, ${\cal H}$, $\Xi\ln G$, $xQ/R^3$, $M-\tilde M$, and
$V$ (in fe), show clear pointwise second-order convergence, except
near the centre. For these variables, the dominant error looks like
$h^2 \hat f_2(u,i)$ for the first few grid points (where $i$ is the
grid index), transitioning to the expected $h^2 f(u,x)$ at larger
$i$. This still implies convergence in any function norm.

We note in passing that at $u=0$, where $\psi$, $\chi$ and either $R$
or $G$ or both are still exact, $\Xi\psi$ and $\Xi\chi$ converge to
third order.

The results for the six combinations of initial Bondi gauge or
initial affine gauge with the eR, eG and fe formulations
are broadly similar.

In order to test the Raychaudhuri equation $E_{xx}=0$, we have also
discretized
\begin{eqnarray}
E_{xx(1)}&:=&G V_{,x}, \\
E_{xx(2)}&:=&-4\pi R(\psi_{,x}^2+\chi_{,x}^2),
\end{eqnarray}
using a different method from elsewhere in the code. This means in
the eR or eG formulations, where $G$ or $R$,
respectively, are found by solving $E_{xx}=0$, we have a non-trivial
test of the discretization of that process. In the fe
formulation, we have a test of the discretization of the solution of
$E_{xx}=0$ for either $R$ or $G$ in the initial data, combined with a
test of how the subsequent free evolution error leads to increasing
violation of the constraint $E_{xx}=0$. We find that
$E_{xx(1)}-E_{xx(2)}=0$ converges to second order, except at the
boundaries, where it is only first-order. This is because there we use
second-order accurate but one-sided finite differences in evaluating
the constraint, and therefore in self-convergence testing at the
boundaries we compare a one-sided derivatives at the lower resolution
with a centred derivative at the higher resolution.

For all variables, we have also taken the three differences of the
numerical solutions in the three formulations, and again we find
second-order convergence, with errors of the order of magnitude of the
self-convergence errors in each formulation. In other words, all three
formulations converge to the same continuum solution, as expected.

%%%%%%%%%%%%%%%%%%%%%%%%%%%%%%%%%%%%%%%%%%%%%%%%%%%%%%%%%%%%%%%%%%%%%%%%%%%%%

\subsection{Evolving through the event horizon and singularity excision}

%%%%%%%%%%%%%%%%%%%%%%%%%%%%%%%%%%%%%%%%%%%%%%%%%%%%%%%%%%%%%%%%%%%%%%%%%%%%%

We have further evolved the initial data that we already used for
convergence testing through the horizon, obviously in eG and fe
only. We lose pointwise convergence there, but the solutions in
eG and fe remain qualitatively alike. On each time slice beyond the event
horizon, $R$ increases from zero to a maximum that marks the apparent
horizon, then decreases back to zero at some $x=x_s(u)$ which denotes
the spacelike curvature singularity, with $dx_s/du<0$. We excise grid
points with $x>x_s(u)$, which are no longer physical. The code remains
stable and appears accurate up to within very few grid points of the
singularity.

As $x\to x_s$, $R_{,x}$ and $G$ diverge. We therefore also plot
against $u$ and $\lambda$. We find that $R$ then decreases essentially
linearly with $\lambda$ everywhere inside the apparent horizon until
we lose resolution near the singularity.

Inside the apparent horizon, ${\cal M}\simeq 0.21$ and $Q\simeq 0.023$
appear essentially constant, compared to $R$. If ${\cal M}$ and $Q$
were exactly constant, the spacetime would be locally isometric
to Reissner-Nordstr\"om spacetime, and with $Q\ne 0$ the singularity
at $R=0$ should then be timelike, and preceded by a Cauchy
horizon. However, where we excise, $R$ is still a significant fraction
of the event horizon radius $r_+:={\cal M}+\sqrt{{\cal M}^2-Q^2}\simeq
2 {\cal M}$, while the Cauchy horizon radius $r_-:={\cal
  M}-\sqrt{{\cal M}^2-Q^2}\simeq Q^2/(2{\cal M})$ is smaller by a
factor of about $10^{-4}$, so we lack resolution for seeing
the Cauchy horizon.

Clearly, our code is not yet very suitable for exploring aspects of
charged black hole interiors such as Cauchy horizons and mass
inflation, but that is not its purpose at the moment. Rather, we want
to demonstrate that we can evolve into a region where outgoing null
cones reconverge. 

%%%%%%%%%%%%%%%%%%%%%%%%%%%%%%%%%%%%%%%%%%%%%%%%%%%%%%%%%%%%%%%%%%%%%%%%%%%%%

\subsection{Type-II critical collapse of a spherical charged scalar field}

%%%%%%%%%%%%%%%%%%%%%%%%%%%%%%%%%%%%%%%%%%%%%%%%%%%%%%%%%%%%%%%%%%%%%%%%%%%%%

As our emphasis in this paper is on developing methods for simulating
type-II critical collapse on null cones beyond spherical symmetry,
rather than on charged collapse in spherical symmetry in detail, we
examine only four one-parameter families of initial data. In the first
  of these, $\phi(0,x)$ is given by (\ref{chargedinitialdata}) with
  $x_c=1.5$, $d=0.5$, $\omega=2.0$ and ${\cal A}=0.05$, and we
  initialise the radial gauge with $G(0,x)=1$. We set the coupling
  constant to $q=1/\sqrt{4\pi}\simeq 0.2821$. The second family is as
  the first except that we initialise the radial gauge with
  $R(0,x)=x$. As already $R_{,x}(0,x)\simeq 1$ in the first family,
  these two families do not differ qualitatively. The third family is
  again like the first, except that we set $\omega=0$, so that the
  scalar field is real and $Q=0$. The fourth family is again like
  the first, except that we set $q=10/\sqrt{4\pi}\simeq 2.821$, so
  that $Q(0,x)$ is 10 times as large.

During the evolution, we use sdn gauge, given in (\ref{sdnshift}),
with the parameter $x_0$ adjusted so that $x=x_0$ is approximately the
past light cone of the accumulation point of echos in near-critical
solutions.

Numerically, we use either the eR or eG formulation. We have not tried
the fe formulation in critical collapse, as we expect the slow drift
away from the regularity condition $R_{,x}=G$ at the centre to affect
the correctness of near-critical evolutions. As for numerical
parameters, we use a uniform grid in $x$ with $x_\text{max}=5$ and
$N_x=250$ grid points, and the timestep (\ref{shiftandRtres}) with
$C_2=0.1$ and $C_3=0.001$.

For the $G=1$ initial data, we chose $x_0=3.17$ in both the eR and eG
formulation. We found $p_*\simeq 0.91$. The initial mass and charge at
$p\simeq p_*$ are $M=0.151$ and $Q\simeq 0.021$.

{For the $R=x$ initial data, we chose $x_0=3.107$ in the eR
formulation and $x_0=3.103$ in the eG formulation.} We found
$p_*\simeq 1.01$. The initial mass and charge at $p\simeq p_*$ are
$M=0.173$ and $Q\simeq 0.026$.  

The results for the two families of initial data are qualitatively
similar, as one would expect. The results for the eR and eG
formulations, for the same initial data, are quantitatively very
similar, as one would expect from the convergence tests. In the
following, for conciseness we present results only for the $G=1$
family of initial data, evolved with the eG formulation. 

{

As a proxy for the black hole Hawking mass and charge, which are
strictly speaking defined at future null infinity, we evaluate $M$ and $Q$
on the first symmetry 2-sphere where the Hawking compactness takes
the value $C(u,x)=C_\text{thr}$. (``First'' here means the smallest
such $u$.) In the results presented here in the eG
formulation, we choose the obvious value $C_\text{thr}=1$, but this is
not possible in the eR formulation. We discuss the role of the choice
of $C_\text{thr}$ in Sec.~\ref{sec:collapsecriterion}.}

%%%%%%%%%%%%%%%%%%%%%%%%%%%%%%%%%%%%%%%%%%%%%%%%%%%%%%%%%%%%%%%%%%%%%%%%%%%%%
\begin{figure}
\includegraphics[width=\linewidth]{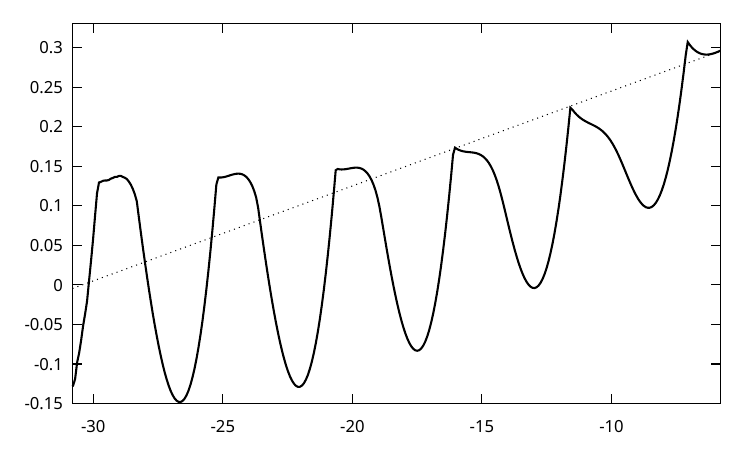} 
\caption{Maximum curvature fine-structure for the $G=1$ family of
  initial data, in the eG formulation: we plot $-1/2\ln|\max
  T|-\gamma\ln(1-p/p_*)+A$ against $\ln(1-p/p_*)$. We use the
  approximate theoretical value $\gamma=0.374$, and the
  family-dependent offset $A=-0.23$ has been fitted by eye so that the
  average of the fine-structure is approximately zero at high
  fine-tuning to the threshold of collapse,(that is, in the limit
  $|p-p_*|\to 0$). The range of $\ln(1-p/p_*)$ on the horizontal axes
  has been chosen as in Fig.~1 of Paper~II for comparison. (As in
  Paper~II, we have suppressed axis labels to make figures larger.)
  The thin dotted line has been fitted by eye to the first two peaks
  from the right and has slope $\delta\gamma=0.012$, meaning that the
  curvature critical exponent is slightly larger at low fine-tuning}.
\label{charged_iG_eG_finestructure_T}
\end{figure}
%%%%%%%%%%%%%%%%%%%%%%%%%%%%%%%%%%%%%%%%%%%%%%%%%%%%%%%%%%%%%%%%%%%%%%%%%%%%%

%%%%%%%%%%%%%%%%%%%%%%%%%%%%%%%%%%%%%%%%%%%%%%%%%%%%%%%%%%%%%%%%%%%%%%%%%%%%%
\begin{figure}
\includegraphics[width=\linewidth]{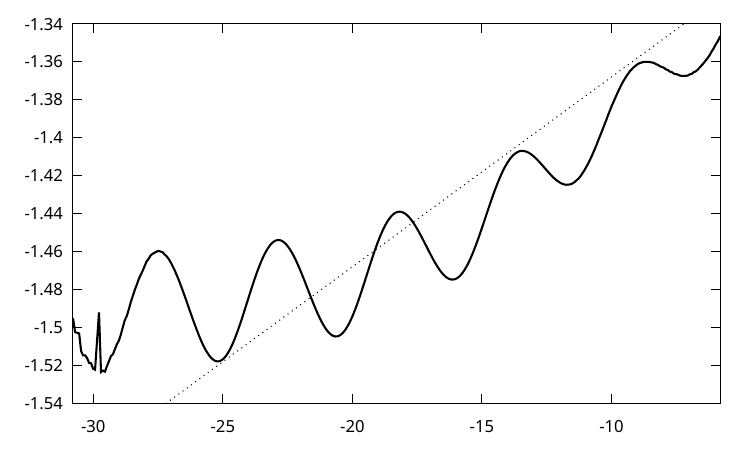} 
\caption{Black hole mass fine-structure for the $G=1$ family of
  initial data, in the eG formulation, with $C_\text{thr}=1$:
  we plot $\ln M-\gamma\ln(p/p_*-1)+A$ against
  $\ln(p/p_*-1)$. $\gamma$ and $A$ and the horizontal range are as for
  the curvature scaling plot in
  Fig.~\ref{charged_iG_eG_finestructure_T}. Compare also Fig.~2 of
  Paper~II. The thin dotted line has been fitted by eye to the first
  two peaks from the right. It has slope $\delta\gamma=0.010$, meaning
  that the mass critical exponent is slightly larger at low
  fine-tuning, in line with that for the curvature, to within a
  plausible fitting error.}
\label{charged_iG_eG_finestructure_M}
\end{figure}
%%%%%%%%%%%%%%%%%%%%%%%%%%%%%%%%%%%%%%%%%%%%%%%%%%%%%%%%%%%%%%%%%%%%%%%%%%%%%

%%%%%%%%%%%%%%%%%%%%%%%%%%%%%%%%%%%%%%%%%%%%%%%%%%%%%%%%%%%%%%%%%%%%%%%%%%%%%
\begin{figure}
\includegraphics[width=\linewidth]{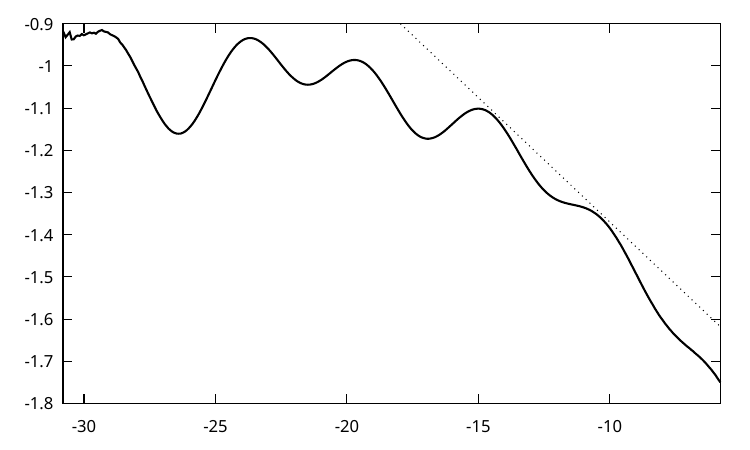} 
\caption{Black hole charge fine-structure for the $G=1$ family of
  initial data, in the eG formulation, with $C_\text{thr}=1$:
  we plot $\ln Q-\mu\ln(p/p_*-1)+A$ against
  $\ln(p/p_*-1)$. We have used the approximate expected critical
  exponent $\mu=0.883$, and $A$ as in the curvature scaling plot in
  Fig.~\ref{charged_iG_eG_finestructure_T}. The thin dotted line has
  been fitted by eye to the first two peaks from right. It has slope
  $\delta\mu=-0.059$, meaning that the charge critical exponent is
  smaller at low fine-tuning. It is plausible from this figure that at
  high fine-tuning the expected value of $\mu$ is achieved. The fine-structure
  has approximately the same period as for the curvature and mass, as
  one expects.}
\label{charged_iG_eG_finestructure_Q}
\end{figure}
%%%%%%%%%%%%%%%%%%%%%%%%%%%%%%%%%%%%%%%%%%%%%%%%%%%%%%%%%%%%%%%%%%%%%%%%%%%%%

%%%%%%%%%%%%%%%%%%%%%%%%%%%%%%%%%%%%%%%%%%%%%%%%%%%%%%%%%%%%%%%%%%%%%%%%%%%%%
\begin{figure}
\includegraphics[width=\linewidth]{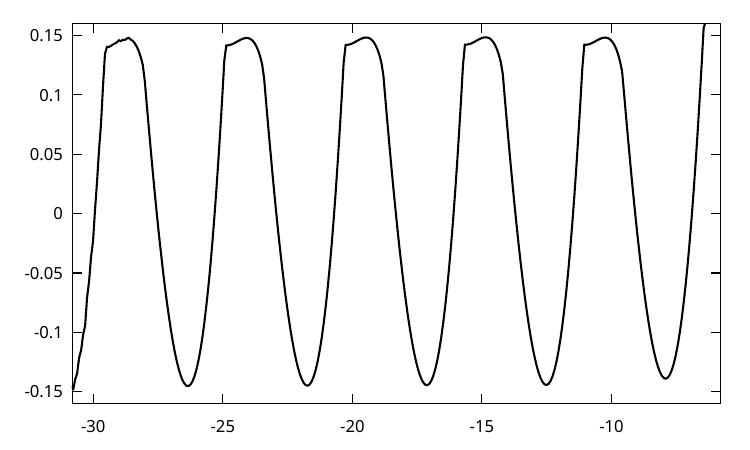} 
\caption{Maximum curvature fine-structure for the $G=1$ family of {\em
    real} initial data, in the eG
  formulation. We set $A=-0.58$, otherwise as in
  Fig.~\ref{charged_iG_eG_finestructure_T}.}
\label{real_iG_eG_finestructure_T}
\end{figure}
%%%%%%%%%%%%%%%%%%%%%%%%%%%%%%%%%%%%%%%%%%%%%%%%%%%%%%%%%%%%%%%%%%%%%%%%%%%%%

%%%%%%%%%%%%%%%%%%%%%%%%%%%%%%%%%%%%%%%%%%%%%%%%%%%%%%%%%%%%%%%%%%%%%%%%%%%%%
\begin{figure}
\includegraphics[width=\linewidth]{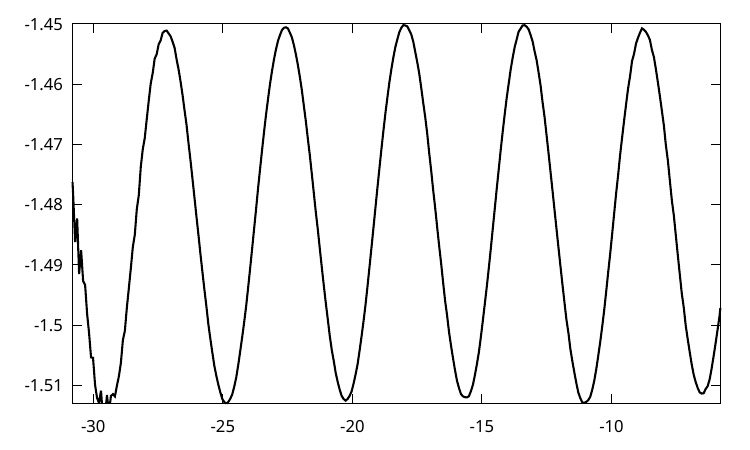} 
\caption{Black hole mass fine-structure for the $G=1$ family of real
  initial data, in the eG formulation, {with
    $C_\text{thr}=1$}. We set $A=-0.58$, otherwise as in
  Fig.~\ref{charged_iG_eG_finestructure_M}.}
\label{real_iG_eG_finestructure_M}
\end{figure}
%%%%%%%%%%%%%%%%%%%%%%%%%%%%%%%%%%%%%%%%%%%%%%%%%%%%%%%%%%%%%%%%%%%%%%%%%%%%%

The fine-structure of critical scaling on the dispersal side of the
threshold of collapse is shown in
Fig.~\ref{charged_iG_eG_finestructure_T} for the curvature invariant
$T$. On the collapse side, Figs.~\ref{charged_iG_eG_finestructure_M}
and \ref{charged_iG_eG_finestructure_Q} show the fine-structure for
$M$ and $Q$ of the first (approximate) marginally outer-trapped
  surface (from now on, MOTS and FMOTS for the first MOTS). What we
actually plot is $-1/2\ln T-\gamma\ln(1-p/p_*)+A$ in
Fig.~\ref{charged_iG_eG_finestructure_T}, $\ln M-\gamma\ln(1-p/p_*)+A$
in Fig.~\ref{charged_iG_eG_finestructure_M}, and $\ln
Q-\mu\ln(1-p/p_*)+A$ in Fig.~\ref{charged_iG_eG_finestructure_Q}. The
family-dependent offset $A$ has been fitted so the average of $-1/2\ln
T-\gamma\ln(1-p/p_*)+A$ is roughly zero at high fine-tuning. $\gamma$
and $\mu$ have been set to their theoretically predicted approximate
values $\gamma=0.374$ and $\mu=0.883$ \cite{GundlachMartin96}.

%%%%%%%%%%%%%%%%%%%%%%%%%%%%%%%%%%%%%%%%%%%%%%%%%%%%%%%%%%%%%%%%%%%%%%%%%%%%

\subsection{The effect of charge on the critical scaling}

%%%%%%%%%%%%%%%%%%%%%%%%%%%%%%%%%%%%%%%%%%%%%%%%%%%%%%%%%%%%%%%%%%%%%%%%%%%%

At high fine-tuning, the fine-structures are periodic, thus confirming
the values of $\gamma$ and $\mu$ from perturbation theory. The fine
structure for $T$ has the same amplitude of about $0.15$ and shape as
in Fig.~1 of Paper~II, which showed $T$ for a one-parameter family of
real, spherically symmetric initial data. At low fine-tuning, the
curve has the same periodic structure but rises slightly with
$\ln(1-p/p_*)$. The same observation applies to the periodic fine
structure of $M$. It has an amplitude $\simeq 0.03$. From
Fig.~\ref{charged_iG_eG_finestructure_Q}, the fine structure in $\ln
Q$ evaluated also appears to be continuous, with amplitude $\simeq
0.15$.

The value of $\mu$ predicted in \cite{GundlachMartin96} was verified
in collapse simulations in \cite{HodPiran97} and \cite{Petryk05}, and
is verified here again by us to be correct sufficiently close to
  the threshold of collapse. Figures in these last two papers showed
a power law with a wiggle, but did not show the wiggle clearly. Hence
our Fig.~\ref{charged_iG_eG_finestructure_Q} is the first plot of the
fine structure of the charge scaling law for any physical system, here
the spherical charged scalar field.

At low fine-tuning, we see deviations from the expected scaling
laws. Heuristically, these can be described as modifications of the
critical exponents $\gamma$ and $\mu$ (as we did for the critical
exponents for the mass and angular momentum in \cite{GunBau18}). To
quantify these modifications, we have fitted the thin dotted lines in
Figs.~\ref{charged_iG_eG_finestructure_T},
\ref{charged_iG_eG_finestructure_M} and
Fig.~\ref{charged_iG_eG_finestructure_Q} by eye. These have slopes of
$\delta\gamma=0.012$, $\delta\gamma=0.010$, and $\delta\mu=-0.059$,
respectively. Note the values of $\delta\gamma$ on the dispersion and
collapse sides are roughly the same, as one would expect, but the
fitting by eye is somewhat arbitrary. As in \cite{GunBau18}, the small
change in critical exponent is a convenient quantitative way of
describing the fine structure at low fine-tuning, but we do not have a
quantitative explanation for it.

For comparison, we have also fine-tuned a family of real, and
therefore uncharged, initial data with $G=1$, differing from our
charged data only by the choice $\omega=0$. We also set ${\cal
  A}=0.075$ in order to make $p_*\simeq 1$. For the eG formulation, we
chose $x_0=3.525$ and found $p_*\simeq 0.95$. For the eR formulation,
we chose $x_0=3.530$ and found $p_*\simeq 0.94$. The fine-structure
for $T$ and $M$ (in the eG formulation, {with $C_\text{thr}=1$}) is
given in Figs.~\ref{real_iG_eG_finestructure_T} and
\ref{real_iG_eG_finestructure_M}. Here we can read off more accurately
that the amplitude of the fine structure for $M$ is $\simeq 0.032$. As
already noted in Paper~II, this differs significantly from the values
for the amplitude given in the literature. It is remarkable how
periodic both fine-structures are already at low fine-tuning, in
contrast to the charged case, where $|Q|/M$ has to be quite small
before the critical exponents agree with their values calculated in
perturbation theory in \cite{GundlachMartin96}. That calculation
considered charge as a linear perturbation on an uncharged background
critical solution (the Choptuik solution).  From
Figs.~\ref{charged_iG_eG_finestructure_T},
\ref{charged_iG_eG_finestructure_M} and
\ref{charged_iG_eG_finestructure_Q}, this seems to be accurate for
$\ln|1-p/p_*|\lesssim -20$. This corresponds to $|Q|/M\lesssim
10^{-4}$ for the black hole.

\begin{figure}
\includegraphics[width=\linewidth]{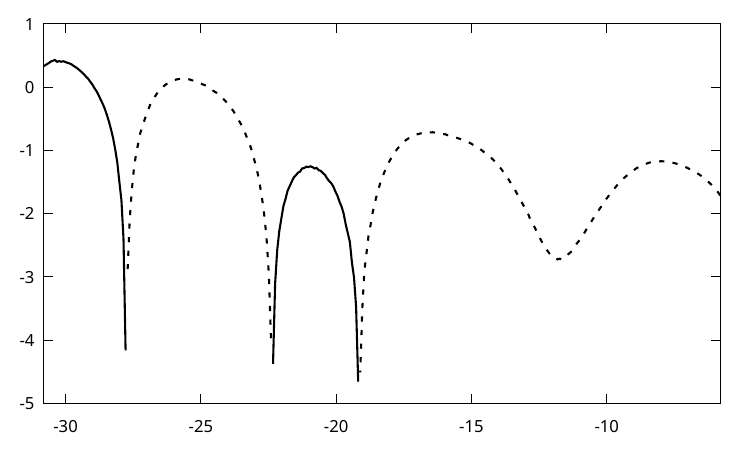} 
\caption{\new Charge fine-structure for a different family of complex
  Gaussian initial data (in the eG formulation, with
  $C_\text{thr}=1$.) The initial scalar field profile is the same, and
  again $G=1$ in the initial data, but
  $q$ is larger by a factor of 10, compared with the family
  represented in Fig.~\ref{charged_iG_eG_finestructure_Q}. For
  plotting, we set $A=-0.23$, otherwise as in
  Fig.~\ref{charged_iG_eG_finestructure_Q}. The charge is positive
  where the curve is solid and negative where the curve is
  dashed. (The kinks in $\ln |Q|$ are where $Q$ goes through zero.)
  This is qualitatively different from
  Fig.~\ref{charged_iG_eG_finestructure_Q}, where the charge is always
  positive.}
\label{L_Q}
\end{figure}

So far we have presented results only for a single
one-parameter family of charged initial data, which tells us nothing
about universality. Both the mass and charge critical exponents appear
to agree with their predicted values, however, which is indirect
evidence of their universality. We also find that the fine-structure
of the mass scaling law looks very similar in the charged and
uncharged families. {\new By considering a more highly charged family, we have
found initial evidence that the fine-structure of the charge scaling
law may not be universal.

We consider again complex Gaussian initial data with $x_c=1.5$,
$d=0.5$, $\omega=2.0$ and ${\cal A}=0.05$, with initial gauge
$G(0,x)=1$. The only difference is that we now set
$q=10/\sqrt{4\pi}\simeq 2.821$, 10 times the value we investigated
before. (As $q$ has dimension of inverse length, this is equivalent to
keeping $q$ fixed and changing all other length scales in the
numerical domain and initial data by factors of 10.) We find
$p_*\simeq 1.157$. Once again we find charge scaling roughly with
$\mu=0.883$, but the sign of the charge now oscillates in
$\ln(p-p_*)$. The charge fine-structure is shown in
Fig.~\ref{L_Q}. The mass scaling, by contrast, is unchanged (and
therefore we do not show it here). We will return to this
non-universality elsewhere.}

%%%%%%%%%%%%%%%%%%%%%%%%%%%%%%%%%%%%%%%%%%%%%%%%%%%%%%%%%%%%%%%%%%%%%%%%%%%%%
%%%%%%%%%%%%%%%%%%%%%%%%%%%%%%%%%%%%%%%%%%%%%%%%%%%%%%%%%%%%%%%%%%%%%%%%%%%%%

%%%%%%%%%%%%%%%%%%%%%%%%%%%%%%%%%%%%%%%%%%%%%%%%%%%%%%%%%%%%%%%%%%%%%%%%%%%%%

\subsection{The role of the collapse criterion}
\label{sec:collapsecriterion}

%%%%%%%%%%%%%%%%%%%%%%%%%%%%%%%%%%%%%%%%%%%%%%%%%%%%%%%%%%%%%%%%%%%%%%%%%%%%%

%%%%%%%%%%%%%%%%%%%%%%%%%%%%%%%%%%%%%%%%%%%%%%%%%%%%%%%%%%%%%%%%%%%%%%%%%%%%%
\begin{figure}
\includegraphics[width=\linewidth]{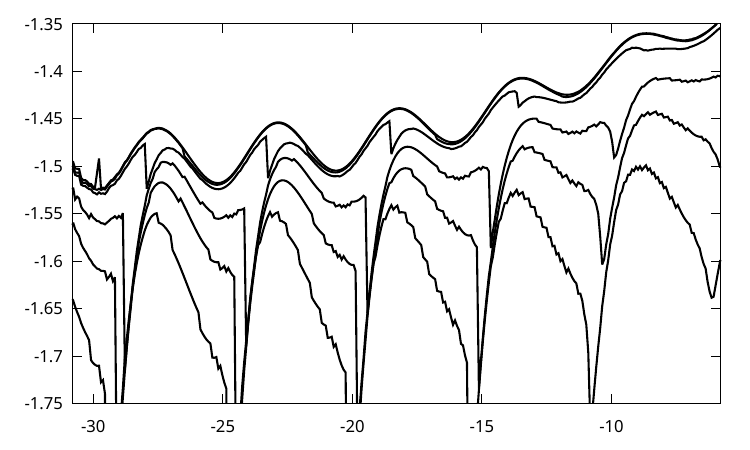} 
\caption{The fine structure of $M$, for the $G=1$ family of initial
  data, in the eG formulation, {but now for different values of
    $C_\text{thr}$. From bottom to top, the curves correspond to
    $C_\text{thr}=0.8$ $0.9$, $0.95$, $0.99$, $0.999$ and $1$, as in
    Fig.~\ref{charged_iG_eG_finestructure_xC}. The
    $C_\text{thr}=0.999$ and $1$ curves are almost indistinguishable,
    but there is a tiny jump once per period in the
    $C_\text{thr}=0.999$ curve. The $C_\text{thr}=1$ curve is the
  same as in Fig.~\ref{charged_iG_eG_finestructure_M}.}}
\label{charged_iG_eG_finestructure_M_allCthr}
\end{figure}
%%%%%%%%%%%%%%%%%%%%%%%%%%%%%%%%%%%%%%%%%%%%%%%%%%%%%%%%%%%%%%%%%%%%%%%%%%%%%

%%%%%%%%%%%%%%%%%%%%%%%%%%%%%%%%%%%%%%%%%%%%%%%%%%%%%%%%%%%%%%%%%%%%%%%%%%%%%
\begin{figure}
\includegraphics[width=\linewidth]{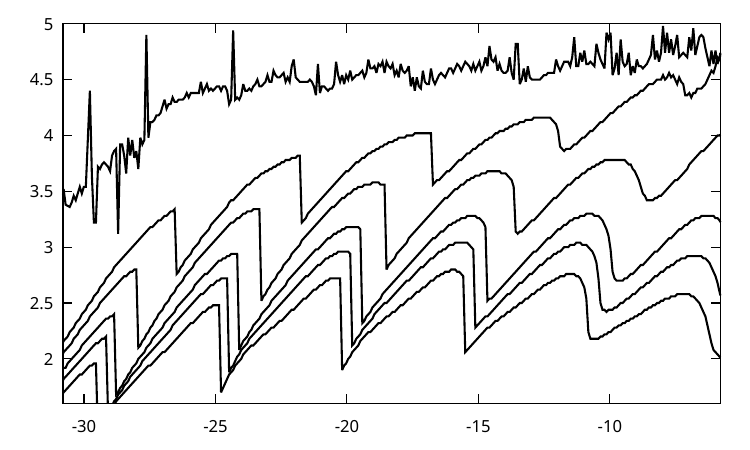} 
\caption{Location $x=x_C$ of the symmetry sphere $(u_C,x_C)$ where we
  evaluate $M$ and $Q$, for the $G=1$ family of initial data, in the
  eG formulation, but now for different values of $C_\text{thr}$. From
  bottom to top, the curves correspond to $C_\text{thr}=0.8$, $0.9$,
  $0.95$, $0.99$, $0.999$ and $1$.}
\label{charged_iG_eG_finestructure_xC}
\end{figure}
%%%%%%%%%%%%%%%%%%%%%%%%%%%%%%%%%%%%%%%%%%%%%%%%%%%%%%%%%%%%%%%%%%%%%%%%%%%%%

In the eG and fe formulations we can choose $C_\text{thr}=1$, but not
in the eR formulation. In Paper~II, for example, where we used the eR
formulation, we used $C_\text{thr}=0.99$ in spherical symmetry and for
mildly non-spherical initial data, and $C_\text{thr}=0.8$ for the
most-nonspherical initial data, and found that the mass
fine-structure was discontinuous, including in spherical symmetry.

To demonstrate the effect of the value of $C_\text{thr}$ on the
measured $M$ and $Q$, Fig.~\ref{charged_iG_eG_finestructure_M_allCthr}
shows the mass fine-structure obtain with different values of
$C_\text{thr}$. From bottom to top, the six curves correspond to
$C_\text{thr}=0.8$, $0.9$, $0.95$, $0.99$, $0.999$ and $1$. The
curves for $C_\text{thr}<1$ are discontinuous once per period, but the
curve for $C_\text{thr}=1$ is continuous. (The jump in the curve for
$C_\text{thr}=0.999$ is just about visible.) Similar jumps are seen in
the fine-structure of $Q$.

Fig.~\ref{charged_iG_eG_finestructure_xC} shows the $x$-location of
the first symmetry 2-sphere $(u_C,x_C)$ where
$C(u_C,x_C)=C_\text{thr}$, for the same values of $C_\text{thr}$. For
each $C_\text{thr}$, this location is discontinuous at the same values
of $\ln(p/p_*-1)$ where $\ln M$ and $\ln Q$ also jump. The jump in
$x_C$ is still large for $C_\text{thr}=0.999$. This suggests that,
with $C_\text{thr}<1$, the jumps in $M$ and $Q$ at periodic values of
$\ln(p/p_*-1)$ are caused by the jumps in $x_C$ at the same values.

The exception is the location $x_C$ of the first occurrence of $C=1$:
it is a very noisy function of $\ln(p/p_*-1)$, but with no obvious
jumps. The noisiness is explained by the difficulty of locating the
first occurrence of $C=1$: in practice, we find the $x$-location where
$C$ goes through 0 at the first timestep where its maximum is larger
than 1. Note that $M$ and $Q$ read off at this noisy location are not
visibly noisy in Figs.~\ref{charged_iG_eG_finestructure_M} and
\ref{charged_iG_eG_finestructure_Q}: we believe this is explained by
the fact that $M$ and $Q$, as functions of $x$ along the curve $C=1$
(the trapping horizon) quickly asymptote to $M_\text{BH}$ and
$Q_\text{BH}$, and so the values of $M$ and $Q$ read off at $x=x_C$
depend only weakly on $x_C$.

{\new

We can understand theoretically why the mass fine-structure {\em in
  spherical symmetry} is continuous for $C_\text{thr}=1$ but
discontinous for $C_\text{thr}<1$. We expect the function $C(u,x;p)$
to be continuous in all its arguments. Nevertheless, the set of points
$(u,x)$ where $C(u,x;p)=C_\text{thr}$ is in general discontinuous in
$p$, a simple example of catastrophe theory. The same is therefore
true for the point in this set with the smallest value of $u$.

However, $C_\text{thr}=1$ in spherical symmetry is an exception to
this generic behaviour, because $C(u,x;p)=1$ is the trapping horizon,
which divides spacetime into a trapped and an untrapped region, so it
cannot jump in $p$.

{\em Beyond spherical symmetry}, a 2-sphere with $C=1$ is not in
general marginally trapped. Therefore we generically expect the mass of the first
coordinate 2-sphere with $C=1$ to jump in $p$. We will test this when
we apply the eG formulation to axisymmetric critical collapse
elsewhere.}

As a practical discovery, we find that plotting the equivalent of
Fig.~\ref{charged_iG_eG_finestructure_xC} just for the data points
obtained in a bisection to $p_*$ allows us to quickly find the
appropriate value of $x_0$: if these curves drop to $x=0$ with
increased fine-tuning, our choice of $x_0$ is too large. If they rise to
$x=x_\text{max}$ then $x_0$ is too small. To find $p_*$ and $x_0$, we
then alternate an automated bisection in $p$ (to machine precision)
with an adjustment of $x_0$ by hand (to about two to four digits). 

%%%%%%%%%%%%%%%%%%%%%%%%%%%%%%%%%%%%%%%%%%%%%%%%%%%%%%%%%%%%%%%%%%%%%%%%%%%%%

\section{Tests on a null rectangle}
\label{sec:nullrectangle}

%%%%%%%%%%%%%%%%%%%%%%%%%%%%%%%%%%%%%%%%%%%%%%%%%%%%%%%%%%%%%%%%%%%%%%%%%%%%%

\subsection{Previous work}

%%%%%%%%%%%%%%%%%%%%%%%%%%%%%%%%%%%%%%%%%%%%%%%%%%%%%%%%%%%%%%%%%%%%%%%%%%%%%

We now test evolutions with null data given on an intersecting
outgoing null cone $u=u_0$, for $v_0\le v\le v_1$ (the ``right
boundary''), and an ingoing null cone $v=v_0$ , for $u_0\le u\le u_1$
(the ``left boundary''). These data determine the solution on the
``null rectangle'' $u_0\le u\le u_1$, $v_0\le v\le v_1$. 

Simulations on null rectangles have been used in
\cite{Gnedin93,BradySmith95,MurReaTan13} to study black hole interiors
in the simpler case of the spherical Einstein-Maxwell-{\em real}
scalar field system. 

By contrast, in \cite{GellesPretorius25}, the spherical
Maxwell-charged scalar system is evolved on a {\em fixed} RN
spacetime, in order to investigate event horizon instabilities. That
work uses double-null coordinates, compactified to both past and
future null infinity (but finite across the past and future event
horizon) as in \cite{MurReaTan13}. It uses the electromagnetic gauge
$\nabla_a(R^{-2}A^a)=0$, or $A_{u,v}+A_{v,u}=0$ in double-null
coordinates, in contrast to our electromagnetic gauge
$A_x=0$, which in double-null coordinates becomes $A_v=0$.  

We are aware of only one previous numerical treatment of the full
Einstein-{\em charged} scalar system in spherical symmetry: in
\cite{BaakeRinne16}, the evolution is on hyperboloidal {\em spacelike}
slices that on the left end are truncated with an excision boundary
inside the event horizon of a RN black hole, and at their right end
extend to future null infinity.

We believe the present paper gives the first algorithm for simulating
the spherical Einstein-Maxwell-charged scalar system in null
coordinates, allowing for simulations both on a null rectangle or with
a regular centre.

%%%%%%%%%%%%%%%%%%%%%%%%%%%%%%%%%%%%%%%%%%%%%%%%%%%%%%%%%%%%%%%%%%%%%%%%%%%%%

\subsection{Physical setup}

%%%%%%%%%%%%%%%%%%%%%%%%%%%%%%%%%%%%%%%%%%%%%%%%%%%%%%%%%%%%%%%%%%%%%%%%%%%%%

We evolve by advancing in $u$, considered as our time coordinate, as
before, with the left boundary identified with $x=0$. We refer to
quantities such as $R_*:=R(u_0,v_0)$ as the ``corner data'' and denote
them by an asterisk. The details of how consistent left and right data
are found are given in Appendix~\ref{appendix:ingoingdata}, and the
discretisation of that algorithm in
Appendix~\ref{appendix:leftnumerical}. Here we only summarise our free
data and gauge choices. On the left boundary, we choose $B=0$ (so that
the left boundary is null), $G=G_*$ (so that $u$ is an
affine parameter on the left boundary), and $A=0$ (an electromagnetic
gauge choice). On the right boundary $u=0$, we choose the $G=G_*$ (so
that $x$ is an affine parameter there). We can freely specify $\psi$
and $\chi$ on the left and right boundaries, the gauge-independent
corner values $R_*$, ${\cal M}_*$ and $Q_*$, and any two out of the
three gauge-dependent corner values $G_*$, $R_{,x*}$ and $R_{,u*}$.

Our algorithm also allows us to fix the right spacetime gauge as
$R_{,x}=R_{,x*}$, but for brevity we do not present results here. When
there is no scalar field on the right boundary, Bondi and affine gauge are
equivalent there anyway. Morever, in our discretisation they are
equivalent without discretisation error.

We can continue the spacetime gauge into the numerical domain with any
choice of $B(u,x)$ (as long as $B=0$ on the left boundary), but again
for brevity we only present double-null gauge $B(u,x)=0$, so that $x$
is an ingoing null coordinate everywhere, and the numerical domain is
the full domain of dependence of the data (a null rectangle). We use
the timestep $\Delta u_0=C_0\Delta x$, with $C_0=0.1$.

We shall restrict to scenarios where $C_*<1$, so that the corner
2-sphere is untrapped. Then $R_{,x*}>0$ gives us $R_{,u*}<0$, and the
ingoing Raychaudhuri equation (\ref{XXR}) gives us $R_{,u}<0$ on the
entire left boundary. We also restrict to the case where the entire right
boundary is untrapped, so that $R_{,x}>0$ there.

%%%%%%%%%%%%%%%%%%%%%%%%%%%%%%%%%%%%%%%%%%%%%%%%%%%%%%%%%%%%%%%%%%%%%%%%%%%%%

\subsection{Prompt collapse}

%%%%%%%%%%%%%%%%%%%%%%%%%%%%%%%%%%%%%%%%%%%%%%%%%%%%%%%%%%%%%%%%%%%%%%%%%%%%%
 
%%%%%%%%%%%%%%%%%%%%%%%%%%%%%%%%%%%%%%%%%%%%%%%%%%%%%%%%%%%%%%%%%%%%%%%%%%%%%
\begin{figure}
\includegraphics[width=0.5\linewidth]{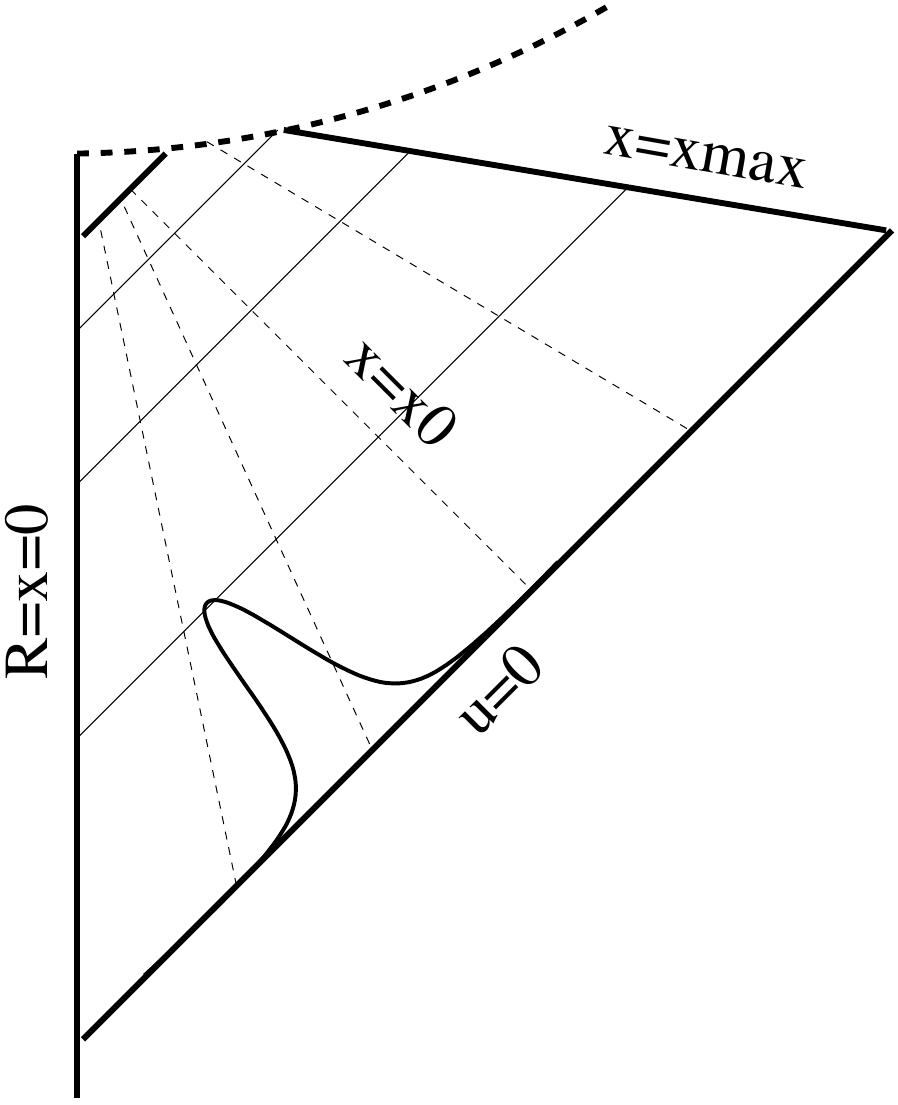} 
\caption{\new Spacetime diagram illustrating collapse from initial data on
  an outgoing null cone, with singularity excision. The regular
  centre and numerical domain are shown as bold lines, the singularity as
  bold dashed, lines of constant $u$ as thin solid, and lines of
  constant $x$ as thin dashed. The lines of constant $x$ illustrate
  sdn gauge, with $x=0$ the timelike regular centre $R=0$, $x=x_0$ ingoing
  null, and $x=x_\text{max}$ future spacelike. Initial data can be
  imposed on the outgoing null cone (with regular centre) $u=0$,
  symbolised here by a bell curve.}
\label{fig:regularcentre}
\end{figure}
%%%%%%%%%%%%%%%%%%%%%%%%%%%%%%%%%%%%%%%%%%%%%%%%%%%%%%%%%%%%%%%%%%%%%%%%%%%%%

%%%%%%%%%%%%%%%%%%%%%%%%%%%%%%%%%%%%%%%%%%%%%%%%%%%%%%%%%%%%%%%%%%%%%%%%%%%%%
\begin{figure}
\includegraphics[width=0.5\linewidth]{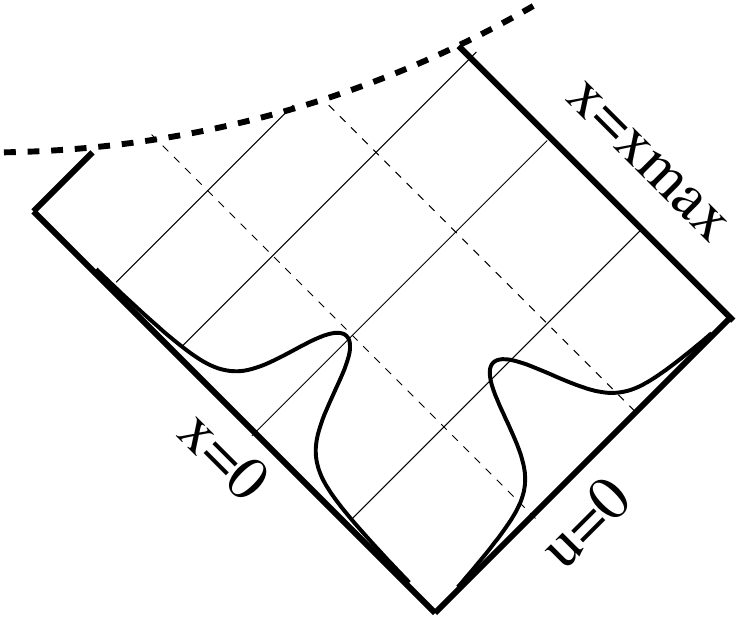} 
\caption{\new Spacetime diagram illustrating the null rectangle
  setup, with initial data imposed on the ingoing null cone $x=0$ and/or the
  outgoing null cone $u=0$. The lines of constant $u$ (thin solid) are
  outgoing null again, as in Fig.~\ref{fig:regularcentre}, but the
  lines of constant $x$ (thin dashed) are all ingoing null,
  illustrating double-null gauge.}
\label{fig:nullrectangle}
\end{figure}
%%%%%%%%%%%%%%%%%%%%%%%%%%%%%%%%%%%%%%%%%%%%%%%%%%%%%%%%%%%%%%%%%%%%%%%%%%%%%

%%%%%%%%%%%%%%%%%%%%%%%%%%%%%%%%%%%%%%%%%%%%%%%%%%%%%%%%%%%%%%%%%%%%%%%%%%%%%
\begin{figure}
\includegraphics[width=0.5\linewidth]{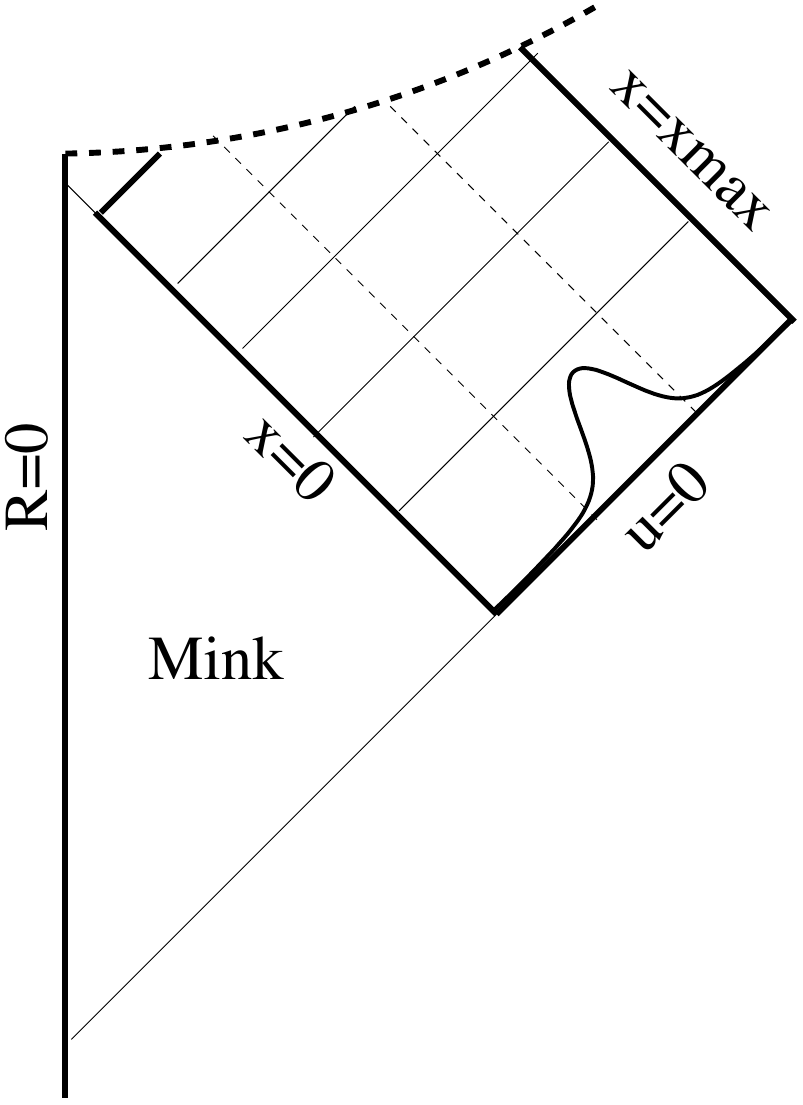} 
\caption{\new Spacetime diagram illustrating what we have called
  ``prompt collapse''. Free initial data are imposed on $u=0$, but the
  data on $x=0$ are now specifically for Minkowski spacetime, so that
  the solution on the numerical domain can be continued into the
  triangle bounded by $R=0$, $x=0$ and $u=0$ (extended as thin solid
  lines) as a part of Minkowski spacetime.}
\label{fig:promptcollapse}
\end{figure}
%%%%%%%%%%%%%%%%%%%%%%%%%%%%%%%%%%%%%%%%%%%%%%%%%%%%%%%%%%%%%%%%%%%%%%%%%%%%%

Consider a null rectangle with $R_*>0$, $R>0$ on the left boundary,
$M_*=Q_*=0$, zero scalar field on the left boundary, and no trapped
surface on the right boundary. Then the triangular region bounded by
the regular centre $R=0$, $u=u_0$ and $v=v_0$ is a piece of Minkowski
spacetime, and $u=u_0$ can be extended to the
left to $R=0$, with vanishing scalar field.

If the data on $u=0$ are sufficiently strong that a MOTS first forms
at $(v_1,u_\text{FMOTS})$ for some $u_0<u_\text{FMOTS}<u_1$, then
there must be an event horizon at some $u=u_\text{EH}$ with
$u_0<u_\text{EH}\le u_\text{FMOTS}$. We have in effect simulated
gravitational collapse from initial data on a null cone with regular
centre, while actually solving the Einstein equations only on a null
rectangle, and thus avoiding the complications of a regular
centre. The null rectangle may then be truncated by a curvature
singularity ``at the top''. {\new Figs.~\ref{fig:regularcentre} and 
\ref{fig:nullrectangle} show gravitational collapse from data on a
single null cone, and from data on two intersecting null cones.
Fig.~\ref{fig:promptcollapse} illustrates the prompt collapse
scenario. In each case, we illustrate excision of a future spacelike
singularity from the numerical domain.}

If, on the other hand, there is no trapped surface in the null
rectangle, formation of a black hole from the regular data on $u=u_0$
is neither proved nor disproved. In particular, we do not expect to
find the black hole threshold in a one-parameter family of initial
data on $u=u_0$ in this setup: $u_\text{FMOTS}$ will become larger than
$u_1$ well above the collapse threshold. Another way of seeing this is
to note that the critical solution in type-II critical collapse is
non-vacuum both inside and outside the past null cone of its
singularity.

As an example of prompt collapse, we set ${\cal M}_*=Q_*=0$, $R_*=2$,
and fix the corner gauge as $G_*=R_{,x*}=1$. With no scalar field on
the left boundary, $R_{,u}=R_{,u*}=-1/2$ there, so the left boundary
intersects $R=0$ at $u=4$. On the right boundary we set $\phi$ to be a
Gaussian with centre at $x_c=0.5$, width $d=0.1$, frequency
$\omega=50$ and amplitude $9.5\cdot 10^{-3}$. We set $q=50$ and
$x_\text{max}=1$.  From now on, without loss of generality, we let
$u_0=0$ and $v_0=v$, and we identify $v=0$ with $x=0$. The event
horizon forms just after $u=3.3$. In the eG and fe formulations we
evolve to $u=3.9$, just before singularity excision starts, and in the
eR formulation at $u=3.3$, just before the event horizon forms.

We evolve with $N_x=100\cdot 2^n$ grid points, with $n=0...4$.  In all
three formulations we find perfect second-order pointwise
self-convergence of $R$ and $\ln G$ already at the lowest resolution,
without any of the complications caused by using expansions at a
regular centre. Perfect second-order self-convergence persists through
the event horizon and until excision starts, and even during excision
breaks down only near the excision boundary. The pairwise differences
between the three formulations similarly show perfect second-order
pointwise convergence.

%%%%%%%%%%%%%%%%%%%%%%%%%%%%%%%%%%%%%%%%%%%%%%%%%%%%%%%%%%%%%%%%%%%%%%%%%%%%%

\subsection{Schwarzschild and extremal Reissner-Nordstr\"om}

%%%%%%%%%%%%%%%%%%%%%%%%%%%%%%%%%%%%%%%%%%%%%%%%%%%%%%%%%%%%%%%%%%%%%%%%%%%%%

In our gauge, the vacuum Schwarzschild solution is already a
non-trivial test, but with an exact solution available, see
Appendix~\ref{appendix:RN}. With the choice $G_*=R_{,x*}=1$, affine
and Bondi right gauge coincide, with $G=R_{,x}=1$ on the entire right
boundary. We set $M_*=0.5$, so that the horizon radius is $r_+=1$, and
$R_*=2$, so that $C_*=1/2$ and $R_{,u}=R_{,u*}=-1/4$ on the entire
left boundary. We set $x_\text{max}=1$. In the eR formulation we stop
the evolution at $u=3.9$, just before the event horizon $R=1$ is
crossed at $u=4$. In the eG and fe formulations, singularity excision
starts at $u\simeq 4.3$, and we stop the evolution at $u=6$, when all
grid points at $x\gtrsim 0.1$ are excised. We again evolve with
$N_x=100\cdot 2^n$, with $n=0...4$. We see perfect self-convergence
and convergence between resolutions to the end of the simulations,
that is close to the horizon in eR and well into the excision regime
in eG and fe. We also see perfect convergence of all formulations
against the exact solution (we have coded up the exact solution only
outside the horizon).

We continue to extremal RN. We set ${\cal M}_*=Q_*=1$, so that the
horizon radius is again $r_+=1$, and we again set $R_*=2$. $M$ is now
not constant, and in particular at the corner $M_*=3/4$, so that
$C_*=3/4$. We again fix the gauge as $G_*=R_{,x*}=1$. This gives us
$R_{,u}=R_{,u*}=-1/8$ on the left boundary, so that the event horizon is
crossed at $u=8$, and the left boundary runs into the 
singularity at $u=16$. We set $x_\text{max}=1$. 

On the horizon $R_{,x}=0$ but, in contrast to the subextremal case,
inside the black hole $R_{,x}>0$ again, and the singularity $R=0$ is
now timelike. This means that our time slices of constant $u$ run into
the singularity at their left edge first. Singularity excision would
not make sense, as boundary conditions would have to be imposed at the
timelike singularity.

We evolve with $N_x=100\cdot 2^n$, with $n=0...6$, up to $u=7.9$ in
the eR formulation, and up to $u=14$ in the eG and fe formulations. We
see perfect self-convergence and convergence between formulations in
the highest resolutions, but convergence in the lower resolutions
begins to break down between $u=13$ and $u=14$, due to the appearance
of sharp gradients in all variables near the left boundary.

%%%%%%%%%%%%%%%%%%%%%%%%%%%%%%%%%%%%%%%%%%%%%%%%%%%%%%%%%%%%%%%%%%%%%%%%%%%%%

\subsection{Perturbed extremal Reissner-Nordstr\"om}

%%%%%%%%%%%%%%%%%%%%%%%%%%%%%%%%%%%%%%%%%%%%%%%%%%%%%%%%%%%%%%%%%%%%%%%%%%%%%

To simulate extremal RN perturbed by charged matter entering through
the {\em right} boundary, we set again ${\cal M}_*=Q_*=1$, $R_*=2$,
$G_*=R_{,x*}=1$, $x_\text{max}=1$, and we evolve to $u=14$, this time
only in the eG and fe formulations as we want to cross the horizon.

We set $\phi=0$ on the left boundary, but on the right boundary we set
$\phi$ to a complex Gaussian of the form (\ref{chargedinitialdata})
with centre at $x_c=0.4$, width
$d=0.1$, complex phase frequency $\omega=30$ and amplitude $5\cdot
10^{-4}$. We set $q=100$. Then at $u=0$ the augmented mass rises from
${\cal M}_*=1$ at $x=0$ to ${\cal M}\simeq 1.0032$ at $x_\text{max}=1$
(the outer boundary), and the charge from $Q_*=1$ to $Q\simeq
1.0068$. One can interpret this as trying to overcharge a pre-existing
static eRN black hole with a burst of scalar field radiation coming in from
past null infinity at finite advanced time.

Not unexpectedly, this particular attempt fails. Indeed for all
$u>1.7$, already well before the putative horizon is crossed,
$Q\le{\cal M}\le 1$ everywhere on the domain $0\le x\le 1$ (with
equality only at the left boundary). The event horizon is crossed at
$u\simeq 7.4$, that is, earlier than in the unperturbed eRN solution
and has radius $R\simeq 1.08$ (and hence Hawking mass $M\simeq 0.54$),
augmented mass ${\cal M}\simeq 0.997$ and charge $Q\simeq 0.993$, so
the final black hole is slightly subcritical.

\begin{figure}
\includegraphics[width=1.0\linewidth]{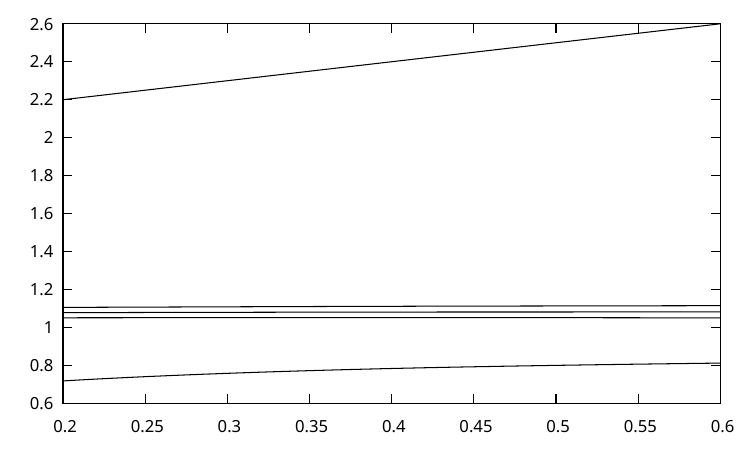} 
\includegraphics[width=1.0\linewidth]{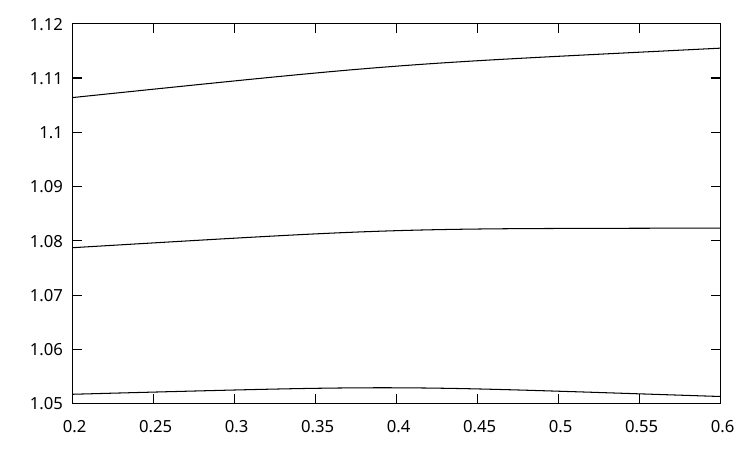} 
\caption{\new Plot of the area radius $R(u,x,)$ in the eRN solution perturbed by a complex Gaussian on the
  right boundary $u=0$. We plot against $x$, at $u=0$, $7.2$, $7.4$,
  $7.6$ and $14$ (from top to bottom). The lower plot shows only
  $u=7.2$, $7.4$ and $7.6$ for clarity, demonstrating that the event
  horizon is crossed at $u\simeq 7.4$.}
\label{fig:peRN_iG_eG_sRtres_1600_R}
\end{figure}

$R$ on our numerical domain $0\le x\le 1$ levels off in $x$ at
$u\simeq 7.4$, indicating the event horizon, then has a maximum in $x$
between $u\simeq 7.4$ and $u\simeq 8.8$, and then returns to be an
increasing function of $x$ at least up to $u=14$, when we stop the
simulation. This is illustrated in
Fig.~\ref{fig:peRN_iG_eG_sRtres_1600_R}. That we do not see $R_{,x}<0$
for sufficiently large $x$, for all $u>u_\text{horizon}$, is probably
due to the limited size of our numerical domain. We will investigate
this elsewhere.

%%%%%%%%%%%%%%%%%%%%%%%%%%%%%%%%%%%%%%%%%%%%%%%%%%%%%%%%%%%%%%%%%%%%%%%%%%%%%
%%%%%%%%%%%%%%%%%%%%%%%%%%%%%%%%%%%%%%%%%%%%%%%%%%%%%%%%%%%%%%%%%%%%%%%%%%%%%

\begin{figure}
\includegraphics[width=1.0\linewidth]{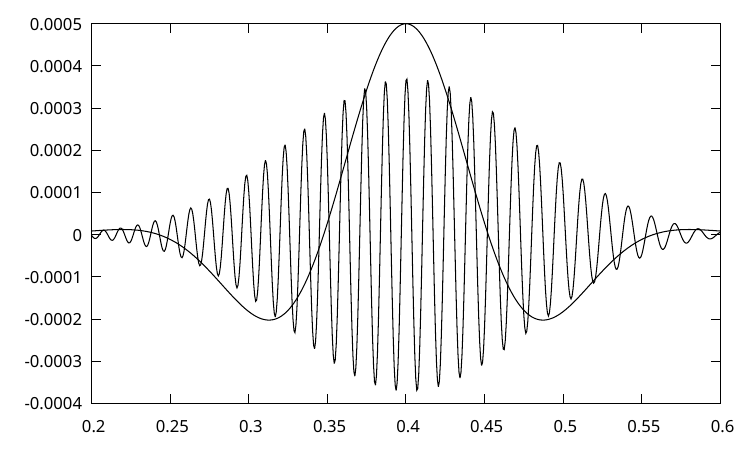} 
\caption{\new Plot of the real part $\psi(u,x,)$ of the complex scalar
  field in the eRN solution perturbed by a complex Gaussian on the
  right boundary $u=0$. We plot against $x$, at $u=0$ (the simpler
  graph) and at $u=14$ (the highly oscillating graph). The imaginary part
  $\chi$ looks similar, with a phase shifting indicating that the
  rapid oscillation is in the complex phase. The numerical domain is
  $0\le x\le 1$, but we plot only the range where $\psi$ is large.}
\label{fig:peRN_iG_eG_sRtres_1600_psi}
\end{figure}

We evolve with $N_x=100\cdot 2^n$, with $n=0...6$. We need the
higher resolutions to see clean self-convergence and convergence
between resolutions. The simulation is more demanding than that of
unperturbed eRN because $\phi$ becomes rapidly oscillating in
$x$. {\new This is illustrated in Fig.~\ref{fig:peRN_iG_eG_sRtres_1600_psi}.}

%%%%%%%%%%%%%%%%%%%%%%%%%%%%%%%%%%%%%%%%%%%%%%%%%%%%%%%%%%%%%%%%%%%%%%%%%%%%%
%%%%%%%%%%%%%%%%%%%%%%%%%%%%%%%%%%%%%%%%%%%%%%%%%%%%%%%%%%%%%%%%%%%%%%%%%%%%%

To simulate extremal RN perturbed by charged matter entering through
the {\em left} boundary, we set once again ${\cal M}_*=Q_*=1$,
$R_*=2$, $G_*=R_{,x*}=1$, $x_\text{max}=1$, and evolve to $u=14$ in
the eG and fe formulations. We now set $\phi=0$ on the right boundary,
while on the left boundary we set $\phi$ to be a complex Gaussian with
centre at $u_c=4$, width $d=0.8$, frequency $\omega=3.75$ and
amplitude $5\cdot 10^{-4}$. (We have scaled $d$ and $\omega$ with
respect to the previous example by factors of $8$ because $R_{,x*}=1$
but $R_{,u*}=-1/8$.) We again set $q=100$. Between $u=3$ and $u=5$,
due to the left data, the augmented mass at $x=0$ drops from ${\cal
  M}_*=1$ to ${\cal M}\simeq 0.9982$ and the charge from $Q_*=1$ to
$Q\simeq 0.9973$, and both $\cal M$ and $Q$ remain almost constant in
$x$ throughout. The horizon is crossed at $u\simeq 7.7$ with
  $R\simeq 1.04$, that is, earlier than in the unperturbed eRN
  solution. Once again we know from $\cal M$ and $Q$ on the horizon
that the black hole is actually slightly sub-extremal, but on our
numerical domain $R$ we do not see $R_{,x}<0$ soon after
the horizon is crossed.

Once again we see good convergence within formulations and between
formulations at the highest resolution. At $u=14$, the complex scalar
field is rapidly oscillating in $x$, with about 16 phase oscillations
between $x=0$ and $x=0.1$, with an exponentially decaying envelope,
and this is only just resolved at the two highest resolutions.

%%%%%%%%%%%%%%%%%%%%%%%%%%%%%%%%%%%%%%%%%%%%%%%%%%%%%%%%%%%%%%%%%%%%%%%%%%%%%

\section{Conclusions}
\label{sec:conclusions}

%%%%%%%%%%%%%%%%%%%%%%%%%%%%%%%%%%%%%%%%%%%%%%%%%%%%%%%%%%%%%%%%%%%%%%%%%%%%%

This series of papers investigates the usefulness of single null
coordinates, in which surfaces of constant retarded time $u$ are null
cones and the coordinate lines of constant $(u,\theta,\varphi)$ are
their generators. The remaining gauge freedom is how points on each
generator are labelled by the radial coordinate $x$.

We have identified three generic formulations: the metric coefficient
$R$ is evolved and the metric coefficient $G$ is found by solving the
(outgoing) Raychaudhuri equation, or the other way around, or both $R$
and $G$ are evolved freely (and the Raychaudhuri equation is imposed
as a constraint only at $u=0$). In these three, any labelling of
points on a null generator by the radial coordinate $x$ is
possible. Dynamically, this radial gauge choice is controlled by
freely choosing $B$. There are also two particular formulations each
tied to a specific gauge: the Bondi formulation where $R=x$, and the
affine formulation where $G=1$. These are maximally constrained, in
the sense that only physical degrees of freedom are evolved.

To our knowledge, there is no clear statement in the literature that
any of the three general formulations can be combined with any choice
of radial gauge. This observation is the main conceptual novelty of
the present paper.

{\new The main motivation for} introducing the eG and fe formulations,
in which $G$ is evolved, is that in them we can evolve through event
and apparent horizons. In Paper~II, both our choice of the eR
formulation and our choice of lsB gauge did not allow us to do this in
axisymmetric scalar collapse. This in turn meant that we could not
bisect arbitrarily close to the collapse threshold in
highly-nonspherical one-parameter families of initial data. {\new In
  the eG and fe formulations, we can now evolve through horizons,
  while still using $B$ to make the coordinate adjust to the
  approximate self-similarity observed in type-II critical collapse.}
We plan to apply our eG formulation with a suitable radial gauge to
this problem.

To present the new formulations pedagogically, and with a focus on
regularity at the central worldline $R=0$, in this paper we have
restricted attention to spherical symmetry. The generalisation beyond
spherical symmetry and/or to higher dimensions is conceptually
straightforward, {\new and will be given elsewhere. As an example in
  spherical symmetry that is also of interest in its own right}, we
have implemented the spherical Einstein-Maxwell-charged scalar
system. We allow for initial data either on a single outgoing null
cone $u=0$ emerging from a regular centre $R=0$, or on two
intersecting null cones $u=u_0$ and $v=v_0$.

In both setups we have shown second-order self-convergence in all
three formulations and second-order convergence between all three
formulations. With a regular centre, the convergence is pointwise
except near the centre, where it still holds in any function norm. The
fact that even the fe formulation converges without any tweaks at the
centre came as a pleasant surprise. With the left boundary an ingoing
null cone, the convergence is perfectly pointwise everywhere.

As a physics testbed of collapse with data on an outgoing null cone,
we have investigated spherical type-II critical collapse of a charged
scalar field. This had previously been investigated theoretically in
\cite{GundlachMartin96}, and numerically in
\cite{HodPiran97,Petryk05}, but here we have presented the first accurate plots
of the universal fine structure in the mass and charge scaling
laws. {\new We have also presented some initial evidence for
  non-universality of the charge scaling fine-structure.

As testbeds for} the setup with initial data on two intersecting
null cones, we have simulated gravitational collapse with a Minkowski
interior, the exact Schwarzschild and extreme Reissner-Nordstr\"om
solutions (which are non-trivial in our gauge), and extreme
Reissner-Nordstr\"om perturbed by a complex scalar field on either
$u=u_0$ or $v=v_0$. 

We plan a study of the {solution space of the spherical
  Einstein-Maxwell-charged scalar system}, and in particular an
investigation of the ``extremal critical collapse conjecture'' for the
spherical Einstein-charged scalar system \cite{KehleUnger24},
see also \cite{KehleUnger22}. We anticipate that both types of
numerical domain will be required for this investigation.

\acknowledgments

Laetitia Martel was supported by an EPSRC Doctoral Training Grant to
the University of Southampton.

%%%%%%%%%%%%%%%%%%%%%%%%%%%%%%%%%%%%%%%%%%%%%%%%%%%%%%%%%%%%%%%%%%%%%%%%%%%%%

\appendix

%%%%%%%%%%%%%%%%%%%%%%%%%%%%%%%%%%%%%%%%%%%%%%%%%%%%%%%%%%%%%%%%%%%%%%%%%%%%%

\section{The Reissner-Nordstr\"om metric in null coordinates}
\label{appendix:RN}

%%%%%%%%%%%%%%%%%%%%%%%%%%%%%%%%%%%%%%%%%%%%%%%%%%%%%%%%%%%%%%%%%%%%%%%%%%%%%

This Appendix presents the RN solution in our variables {and
  gauge choice}. In Schwarzschild coordinates, where $R=r$ is a
coordinate, the RN metric is
\begin{equation}
ds^2=-f\,dt^2+f^{-1}\,dr^2+r^2\,d\Omega^2, 
\end{equation}
where
\begin{equation}
\label{fdef}
f(r):=1-{2{\cal M}_0\over r}+{Q_0^2\over r^2},
\end{equation}
and the real constants ${\cal M}_0>0$ and $Q_0$ are the usual mass and
charge parameters of the RN metric, normalised such that $|Q_0|={\cal
  M}_0$ represents an extremal black hole. Our dynamical variables
${\cal M}$ and $Q$ take the constant values ${\cal M}={\cal M}_0$ and
$Q=Q_0$. The event and Cauchy horizon are at
\begin{equation}
\label{rpm}
r_\pm={\cal M}\pm\sqrt{{\cal M}^2-Q^2}.
\end{equation}
By definition the Hawking mass obeys $r=2M$ on the event
horizon. $M=r_+/2$ at the event horizon is also the irreducible mass
of the black hole. In the special case of an extremal black hole,
$r_+=r_-=2M={\cal M}=Q$.

The tortoise radius $r_*$ is defined, up to addition of a constant, by
\begin{equation}
\label{rstardef}
{dr_*\over dr}:={1\over f(r)}.
\end{equation}
With the outgoing and ingoing Eddington-Finkelstein
coordinates defined as
\begin{equation}
\label{ubarvbardef}
\bar u:=t-r_*, \quad \bar v:=t+r_*,
\end{equation}
the metric in the exterior becomes
\begin{eqnarray}
ds^2&=&-f\,d\bar u\,d\bar v+r^2\,d\Omega^2 \\
\label{RNBondi}
&=&-f\,d\bar u^2-2\,d\bar u\,dr+r^2\,d\Omega^2 \\
&=&-f\,d\bar v^2+2\,d\bar v\,dr+r^2\,d\Omega^2.
\end{eqnarray}
(Both $\bar v$ and $\bar u$ increase to the future. In the interior,
we define $\bar u:=r_*-t$ instead to achieve this.)

Writing (\ref{RNBondi}) in our notation, and choosing our coordinate
$x$ to coincide with $r$, the RN metric in Bondi gauge has the metric
coefficients
\begin{eqnarray}
R&=&x, \\
G&=&1, \\
B&=&{1\over 2}\left(1-{2{\cal M}_0\over x}+{Q_0^2\over
  x^2}\right).
\end{eqnarray}
In the double-null coordinates {$u=\bar u$, $x=\bar v$} we have
\begin{eqnarray}
R&=&r\left({\bar v-\bar u}\over 2\right), \\
\bar G&=&{1\over 2}f(R), \\
B&=&0,
\end{eqnarray}
where $r(r_*)$ is defined as the function inverse of $r_*(r)$, and
$\bar G$ denotes the value of $G$ in the particular double-null
coordinates $\bar u$, $\bar v$.

If, for given constants $\bar u_0$, $\bar v_0$, $u_0$, $v_0$, we now define the
alternative double-null coordinates
\begin{eqnarray}
\label{ufromubar}
u(\bar u)&:=&u_0-a\left(r\left({\bar v_0-\bar u\over
2}\right)-r_0\right), \\
\label{vfromvbar}
v(\bar v)&:=&v_0+b\left(r\left({\bar v-\bar u_0\over
2}\right)-r_0\right),
\end{eqnarray}
where
\begin{equation}
r_0:=r\left({\bar v_0-\bar u_0\over 2}\right), 
\end{equation}
we have $u(\bar u_0)=u_0$, $v(\bar v_0)=v_0$, 
  $R_{,u}(u,v_0)=R_{,u*}:=-1/a$ and $R_{,v}(u_0,v)=(R_{,v})_*:=1/b$.
Inverting (\ref{ufromubar},\ref{vfromvbar}), we obtain
\begin{eqnarray}
\bar u(u)&=&\bar v_0-2r_*\left(r_0+{u-u_0\over -a}\right), \\
\bar v(v)&=&\bar u_0+2r_*\left(r_0+{v-v_0\over b}\right), \\
R(u,v)&=&r\Bigl[r_*\left(r_0+{v-v_0\over b}\right) \nonumber \\ &&
+r_*\left(r_0+{u-u_0\over -a}\right)-r_*(r_0)\Bigr].
\end{eqnarray}
We have 
\begin{eqnarray}
{du\over d\bar u}&=&a\,\bar G(\bar u,\bar v_0), \\
{dv\over d\bar v}&=&b\,\bar G(\bar u_0,\bar v), 
\end{eqnarray}
and therefore
\begin{equation}
G(u,v)={\bar G(\bar v,\bar u)\over 
ab\,\bar G(\bar v_0,\bar u)\,\bar G(\bar v,\bar u_0)},
\end{equation}
and hence $G(u_0,v)=G(u,v_0)=G_*$, where
\begin{equation}
G_*:={1\over ab\, \bar G(\bar u_0,\bar v_0)}.
\end{equation}
This is, in implicit form, the exterior metric of RN in the
coordinates we use in our null-rectangle tests.

%%%%%%%%%%%%%%%%%%%%%%%%%%%%%%%%%%%%%%%%%%%%%%%%%%%%%%%%%%%%%%%%%%%%%%%%%%%%%

\section{Regular centre}
\label{appendix:regularcentre}

%%%%%%%%%%%%%%%%%%%%%%%%%%%%%%%%%%%%%%%%%%%%%%%%%%%%%%%%%%%%%%%%%%%%%%%%%%%%%

The Minkowski metric in Bondi coordinates, with $U$ the retarded time
normalised to be proper time at the centre, is
\begin{equation}
ds^2=-dU^2-2dU\,dR+R^2\,d\Omega^2.
\end{equation}
Now let $U=U(u)$ and $R=R(u,x)$. Our metric
coefficients become
\begin{eqnarray}
G&=&U'R_{,x}, \\
B&=&{U'^2+2U'R_{,u}\over 2U'R_{,x}}, \\
\Xi R&=&-{U'\over 2}, \\
{\cal H}&=&-{U''\over U'}.
\end{eqnarray}
We impose $R(u,0)=0$, thus fixing the regular centre at $x=0$, and we
make $u$ proper time at the centre by setting $U=u$.
This gives us
\begin{eqnarray}
G(u,0)&=&R_{,x}(u,0), \\
\label{BBC}
B(u,0)&=&{1\over 2R_{,x}(u,0)}, \\
\Xi R(u,0)&=&-{1\over 2}, \\
\label{calHBC}
{\cal H}(u,0)&=&0.
\end{eqnarray}
From elementary flatness, these conditions at the centre must also
hold in an arbitary curved spherical spacetime with a regular centre
$R=x=0$ and $u$ proper time at the centre. Beyond spherical symmetry,
in spherical coordinates, they will also hold at the regular worldline
$R=x=0$ as long as it is geodesic, see Appendix~III.E.2 of Paper~I.

%%%%%%%%%%%%%%%%%%%%%%%%%%%%%%%%%%%%%%%%%%%%%%%%%%%%%%%%%%%%%%%%%%%%%%%%%%%%%

\section{Expansions at the centre}
\label{appendix:expansions}

%%%%%%%%%%%%%%%%%%%%%%%%%%%%%%%%%%%%%%%%%%%%%%%%%%%%%%%%%%%%%%%%%%%%%%%%%%%%%

For finding $G$ by integration, given $R$, we need
\begin{eqnarray}
\label{gammaexpansion}
\ln\left({G\over R_{,x}}\right)&=&2\pi (\psi_{(1)}^2+\chi_{(1)}^2)x^2
\nonumber \\ && +{4\pi\over
  3}\Bigl[4(\psi_{(1)}\psi_{(2)}+\chi_{(1)}\chi_{(2)}) \nonumber \\ &&
  -{R_{(2)}\over R_{(1)}}(\psi_{(1)}^2+\psi_{(2)}^2)\Bigr]x^3
\nonumber \\
+O(x^4). 
\end{eqnarray}
Equivalently, for finding $R$ by solving a system of two first-order
ODEs, given $G$, we require
\begin{eqnarray}
\label{Rexpansion}
R&=& G_{(0)} x+{G_{(1)}\over 2}x^2
\nonumber \\ && 
+{1\over 3}\left(G_{(2)}-2G_{(0)}\pi(\psi_{(1)}^2+\chi_{(1)}^2)\right)
x^3+O(x^4), \nonumber \\ \\
V&=& 1-2\pi (\psi_{(1)}^2+\chi_{(1)}^2)x^2
\nonumber \\ && 
+{2\pi\over
  3G_{(0)}}\Bigl[G_{(1)}(\psi_{(1)}^2+\chi_{(1)}^2)
\nonumber \\ && 
-8G_{(0)}(\psi_{(1)}\psi_{(2)}+\chi_{(1)}\chi_{(2)})
\Bigr]x^3 +O(x^4).
\end{eqnarray}
Note that, given the definition of $V:=R_{,x}/G$, to be consistent we
should truncate the expansion of $V$ at one order lower in $x$
than that of $R$, but here we have written out all expansion
coefficients that we know in full given a quadratic fit to
the evolved quantities.  We note in passing that the reverse of
(\ref{Rexpansion}) is
\begin{eqnarray}
\label{Gexpansion}
G&=&R_{(1)}+2R_{(2)}x
+ \left(3R_3+2\pi R_{(1)}(\psi_{(1)}^2+\chi_{(1)}^2)\right) x^2 
\nonumber \\ &&+O(x^3).
\end{eqnarray}

We next have
\begin{eqnarray}
{Q\over {4\pi}q}&=&{R_{(1)}^2\over 3}(\chi_{(1)}\psi_{(0)}-\psi_{(1)}\chi_{(0)})x^3
\nonumber \\ && 
+{R_{(1)}\over 2}\Bigl[R_{(2)}(\psi_{(0)}\chi_{(1)}-\chi_{(0)}\psi_{(1)})
\nonumber \\ && 
+R_{(1)}(\psi_{(0)}\chi_{(2)}-\chi_{(0)}\psi_{(2)})\Bigr]x^4
\nonumber \\ && +O(x^5), \\
{A\over {4\pi} q}&=&{R_{(1)}\over 6}(\chi_{(1)}\psi_{(0)}-\psi_{(1)}\chi_{(0)})x^2
\nonumber \\ && 
+{1\over 6}\Bigl[R_{(2)}(\psi_{(0)}\chi_{(1)}-\chi_{(0)}\psi_{(1)})
\nonumber \\ && 
+R_{(1)}(\psi_{(0)}\chi_{(2)}-\chi_{(0)}\psi_{(2)})\Bigr]x^3
\nonumber \\ && +O(x^4).
\end{eqnarray}
Note that $R_{(3)}$ (or $G_{(2)}$) does not appear to this order.

The expansion of $\Xi R$ is
\begin{eqnarray}
\Xi R&=&-{1\over2}-{\pi\over 3}(\psi_{(1)}^2+\chi_{(1)}^2)x^2
\nonumber \\ && 
-{2\pi\over 3}(\psi_{(1)}\psi_{(2)}+\chi_{(1)}\chi_{(2)})x^3+O(x^4),
\end{eqnarray}
where we have already used the expansion of either $R$ or $G$ above,
but the expansion of $R\Xi R$ takes the simpler form
\begin{equation}
R\,\Xi R=-{G_{(0)}\over2}x-{G_{(1)}\over 4}x^2-{G_{(2)}\over 6}x^3+O(x^4),
\end{equation}
The two are of course equivalent for analytic functions, but we use
the latter in the code. 

As one would expect, we can expand $R\hat\Xi\psi$ to one order higher
than $\hat\Xi\psi$, and similarly for $\hat\Xi\chi$. The latter
expansion is simpler, and is
\begin{eqnarray}
R\,\hat\Xi\psi &=& {\psi_{(1)}\over 2}x+{\psi_{(2)}\over 2}x^2+O(x^3), \\
R\,\hat\Xi\chi &=& {\chi_{(1)}\over 2}x+{\chi_{(2)}\over 2}x^2+O(x^3).
\end{eqnarray}
Again none of the expansion coefficients for $R$ (or $G$)
apear to this order.

Finally, we have 
\begin{eqnarray}
{\cal H}&=&{8\pi \over 3R_{(1)}}(\psi_{(1)}^2+\chi_{(1)}^2)x
\nonumber \\ && 
-{4\pi\over
  3}\Biggl[{R_{(2)}\over R_{(1)}^2} (\psi_{(1)}^2+\chi_{(1)}^2)
\nonumber \\ && 
-{3\over R_{(1)}}(\psi_{(1)}\psi_{(2)}+\chi_{(1)}\chi_{(2)})\Biggr]
x^2
\nonumber \\ &&
+O(x^3).
\end{eqnarray}

%%%%%%%%%%%%%%%%%%%%%%%%%%%%%%%%%%%%%%%%%%%%%%%%%%%%%%%%%%%%%%%%%%%%%%%%%%%%%

\section{Radial null geodesics peeling off the horizon}
\label{appendix:geodesicpeeling}

%%%%%%%%%%%%%%%%%%%%%%%%%%%%%%%%%%%%%%%%%%%%%%%%%%%%%%%%%%%%%%%%%%%%%%%%%%%%%

We consider radial null geodesics in the RN metric. From
(\ref{rstardef}) and (\ref{ubarvbardef}) we have
\begin{equation}
r_{,\bar v}(\bar u,\bar v)={1\over 2}f(r). 
\end{equation}
Linearising about the event horizon $r=r_+$, we have
\begin{equation}
\label{peeling}
r_{,\bar v}\simeq \kappa (r-r_+), \quad \kappa:={1\over 2}f'(r_+),
\end{equation}
where $\kappa$ is the surface gravity. In the subextremal case
$\kappa>0$ the surfaces of constant $u$, and hence the outgoing radial
null geodesics that generate them, peel off the event horizon
exponentially. For an arbitrary $\bar v_0$, we define $\hat u(\bar
u)$ by $\hat u(\bar u):=r(\bar u,\bar v_0)-r_+$, so that the solution of
(\ref{peeling}) can be written as
\begin{equation}
\label{peelingbis}
r(\hat u,\bar v)\simeq r_++\hat u e^{\kappa(\bar v-\bar v_0)}.
\end{equation}
Note that the coordinate $\bar u$ in the metric (\ref{RNBondi})
diverges at the horizon, but by construction $\hat u$ is regular.

A neighbourhood of the event horizon in asymptotically flat charged
collapse at sufficiently late time (large $\bar v$) will be
approximated by the event horizon in RN. As both $\hat u$ and our
numerical coordinate $u$ are regular through the horizon, to leading
order we must have $\hat u\simeq \hat c(u-u_h)$ for some constant
$\hat c>0$, where the horizon is at $u=u_h$. Differentiating
(\ref{peelingbis}) with respect to $u$, we find
\begin{equation}
\Xi R\simeq c u e^{\kappa\bar v}
\end{equation}
near the horizon at late times, where $c$ absorbs $\hat c$ and $\bar
v_0$. 

We have so far defined $\bar v$ only on RN, not on the collapse
spacetime that we are trying to model. We note that on RN $\bar v$
obeys $\bar v=\bar u+2R$. If we assume that for sufficiently early $u$
and sufficiently late $\bar v$ the collapse spacetime is approximated by
the RN exterior, we can define $\bar v=2R$ on
$u=0$ and propagate it with $\Xi\bar v=0$.

%%%%%%%%%%%%%%%%%%%%%%%%%%%%%%%%%%%%%%%%%%%%%%%%%%%%%%%%%%%%%%%%%%%%%%%%%%%%%

\section{Boundary conditions on an ingoing null cone intersecting an
  outgoing one}
\label{appendix:ingoingdata}

%%%%%%%%%%%%%%%%%%%%%%%%%%%%%%%%%%%%%%%%%%%%%%%%%%%%%%%%%%%%%%%%%%%%%%%%%%%%%

We want to impose consistent null data on $x=0$, $u>0$ (the left
boundary) and $u=0$, $x>0$ (the right boundary), intersecting at the
point $u=x=0$ (the corner). On the left boundary we choose the
gauge $B=0$ ($x=0$ is null) and $G_{,x}=0$ ($u$ is an
affine parameter), and the electromagnetic gauge $A=0$. Then
$\Xi=\partial_u$ and ${\cal H}=0$ there. In the following, we write
$R(u,0)=:R_l(u)$ and $dR_l/du=:\dot R_l(u)$, and similarly for all other
quantities on the left boundary. 

The ingoing Raychaudhuri equation {(\ref{EEuu}) with $A=B={\cal
    H}=0$} gives
\begin{equation}
\label{leftReqn}
\ddot R_l+4\pi (\dot\psi_l^2+\dot\chi_l^2)R_l=0.
\end{equation}
Note that $\ddot R_l\le 0$, so if $\dot R_{l*}<0$ at the corner (as we
shall always assume), then $\dot R_l<0$ for all $u\ge 0$. Eq.~(\ref{XiQeqn}) with $A=B=0$ gives
\begin{equation}
\label{leftQeqn}
\dot Q_l={4\pi}qR_l^2(\chi_l\dot\psi_l-\psi_l\dot\chi_l),
\end{equation}
and Eq.~(\ref{Cexpr}) with $B=0$ gives
\begin{equation}
\label{leftRxeqn}
{2\dot R_lR_{,xl}\over G_l}=C_l-1,
\end{equation}
where $R_{,xl}(u):=R_{,x}(u,0)$ and $C_l$ is used as a shorthand for 
\begin{equation}
\label{leftCeqn}
C_l:={2\over R_l}\left({\cal M}_l-{Q_l^2\over 2R_l}\right),
\end{equation}
compare the definitions (\ref{Mdef}) and (\ref{calMdef}).
With $A=B=0$, and using (\ref{leftRxeqn}) to eliminate
$R_{,xl}$, (\ref{XicalMexpr}) becomes
\begin{eqnarray}
\label{leftcalMeqn}
\dot{\cal M}_l&=&-2\pi {R_l^2\over\dot R_l}(C_l-1)(\dot\psi_l^2+\dot\chi_l^2)
\nonumber \\ 
&&+4\pi qQ_lR_l(\chi_l\dot\psi_l-\psi_l\dot\chi_l).
\end{eqnarray}
Note that $\dot {\cal M}_l$ can have either sign. 
Complete boundary data are now obtained as follows.

1) We fix the physical corner data $R_*$, ${\cal M}_*$ and $Q_*$ and
define $C_*$ from them by (\ref{leftCeqn}). We fix two of the
three gauge-dependent corner values $G_*$, $\dot R_{l*}$ and $R_{,x*}$,
with the third one given by the constraint
\begin{equation}
{2\dot R_{l*}R_{,x*}\over G_*}=C_*-1.
\end{equation}

2) We fix the gauge on the right boundary $u=0$ by either setting
$G(x,0)=G_*$ (affine initial gauge) or $R(x,0)=R_*+R_{,x*}\,x$ (Bondi
initial gauge). We fix the gauge on the left boundary by setting
$G_l(u)=G_*$ (affine left gauge). We now fix scalar field data
$\psi_l(u)$, $\chi_l(u)$ on the left boundary and $\psi(0,x)$,
$\chi(0,x)$ on the right boundary. 

For the solution to be sufficiently often differentiable in a
neighbourhood of the corner, we need to impose a sufficient number
of corner conditions on the derivatives of the boundary data. Here we
avoid this complication by assuming that the scalar field data are
bounded away from the corner.

3) We solve (\ref{leftReqn}) as a second-order linear ODE for $R_l$,
starting from $R_*$ and $\dot R_{l*}$, then
(\ref{leftQeqn}) as a
first-order ODE for $Q_l$, starting from $Q_*$, and finally
(\ref{leftcalMeqn}) as a first-order ODE for ${\cal M}_l$,
starting from ${\cal M}_*$. 

4) From (\ref{leftRxeqn}) we read off
\begin{equation}
V_l={C_l-1\over 2\dot R_l}. 
\end{equation}
We now have the data that we need to start up the integration of the
hierarchy equations at $x=0$, namely $Q=Q_l$ for (\ref{Qeqn}), $A=0$
for (\ref{Aeqn}), either $R=R_l$ and $V=V_l$ for
(\ref{Vxeqn},\ref{Rxeqn}) or $G=G_*$ for (\ref{Geqn}), $\Xi R=\dot
R_l$ for (\ref{XiReqn}), $\hat\Xi\psi=\dot\psi_l$ for
(\ref{Xipsieqn}), $\hat\Xi\chi=\dot\chi_l$ for (\ref{Xichieqn}), and
${\cal H}=0$ for (\ref{calHeqn}).

In the simple case where $\psi_l=\chi_l=0$ the boundary ODE
system becomes trivial, giving $Q_l=Q_*$, ${\cal M}_l={\cal M}_*$, and 
$R_l=R_*+\dot R_{l*}\,u$.

%%%%%%%%%%%%%%%%%%%%%%%%%%%%%%%%%%%%%%%%%%%%%%%%%%%%%%%%%%%%%%%%%%%%%%%%%%%%%

\section{Discretization of the left boundary equations}
\label{appendix:leftnumerical}

%%%%%%%%%%%%%%%%%%%%%%%%%%%%%%%%%%%%%%%%%%%%%%%%%%%%%%%%%%%%%%%%%%%%%%%%%%%%%

In the continuum, the data on $x=0$ are automatically compatible with
the evolution equations for $\psi$, $\chi$ and $R$ and/or $G$ on
$x\ge0$. To make the discretised versions compatible as well, rather
than $\psi_l$ and $\chi_l$ we specify $\dot\psi_l$ and $\dot\chi_l$,
together with corner values $\psi_*$ and $\chi_*$. We write
(\ref{Reqn}) in first-order form, and use the same ODE solver that we
use to evolve $\psi$, $\chi$ and $R$ and/or $G$ in $u$ in the bulk to
also evolve $Q_l$, ${\cal M}_l$, $\dot R_l$, $R_l$, $\psi_l$ and
$\chi_l$ on the left boundary. With $A=B=0$ on the left boundary, the
values of $\psi_{,u}$, $\chi_{,u}$, $R_{,u}$ and/or $(\ln G)_{,u}=0$
are then the same, up to round-off error, in the boundary ODE system
and at $x=0$ in the bulk PDE system.

The Runge-Kutta ODE solver that we use for the time integration
assumes that the ODE system is autonomous. We add $u$ as an evolved
variable to the set of left-boundary variables, in order to access the
given functions $\dot\psi_l(u)$ and $\dot\chi_l(u)$ at the correct substep
values of $u$.

%%%%%%%%%%%%%%%%%%%%%%%%%%%%%%%%%%%%%%%%%%%%%%%%%%%%%%%%%%%%%%%%%%%%%%%%%%%%%

\section{Convergence testing}
\label{appendix:convergencetesting}

%%%%%%%%%%%%%%%%%%%%%%%%%%%%%%%%%%%%%%%%%%%%%%%%%%%%%%%%%%%%%%%%%%%%%%%%%%%%%

For any computed variable $f$, we define
\begin{equation}
{\cal E}[f](u,x;h):=\left({h\over h_\text{ref}}\right)^{-2}\left(f(u,x;h)-f(u,x;h/2)\right)
\end{equation}
where $h:=\Delta x=x_\text{max}/N_x$, and we assume $\Delta u$ is
proportional to $h$. For a numerical quantity at resolution $h$, given
that our numerical methods are designed to be second-order accurate,
we expect the Richardson expansion
\begin{equation}
\label{Richardson}
f(u,x;h)= f_0(u,x)+h^2f_2(u,x)+h^3f_3(u,x)+O(h^4)
\end{equation}
to hold, where $f_0(u,x)$ is the (unknown) continuum solution and
$f_i(u,x)$ are (discretisation-dependent, unknown) error terms, and
hence
\begin{equation}
{\cal E}[f](u,x;h)=h_\text{ref}^2 \left({3\over 4}f_2(u,x)+{7\over 8}hf_3(u,x)+O(h^2)\right).
\end{equation}
Hence if ${\cal E}[f](u,x;h)$ for different $h$ are almost equal, we
see pointwise second-order convergence, in the sense that the
numerical error is dominated by the term $h^2f_2$. ${\cal
  E}[f](u,x;h)$ evaluated at any $h$ then provides an estimate of the
actual error at the reference resolution $h=h_\text{ref}$.

We can also test the convergence (to zero) of $f:=M-\tilde M$, or of
the difference $f$ of any variable computed in two different
formulations. We then plot
\begin{equation}
{\cal E}[f](u,x;h):=\left({h\over h_\text{ref}}\right)^{-2}f(u,x;h),
\end{equation}
which, as $f_0(u,x)=0$, we expect to take values
\begin{equation}
{\cal E}[f](u,x;h)=h_\text{ref}^2 \left(f_2(u,x)+hf_3(u,x)+O(h^2)\right).
\end{equation}

%%%%%%%%%%%%%%%%%%%%%%%%%%%%%%%%%%%%%%%%%%%%%%%%%%%%%%%%%%%%%%%%%%%%%%%%%%%%%%%

%%%%%%%%%%%%%%%%%%%%%%%%%%%%%%%%%%%%%%%%%%%%%%%%%%%%%%%%%%%%%%%%%%%%%%%%%%%%%

\end{document}